\documentclass[11pt]{article}
\usepackage{amsfonts}
\usepackage{amssymb}
\usepackage{epsfig}
\usepackage{dcolumn}

\advance \textheight by \topskip
\usepackage{amsmath,dsfont}
\usepackage[english]{babel}

\makeatletter
\@addtoreset{equation}{section}
    \def\theequation{\thesection.\arabic{equation}}
\makeatother

\def\Z{\mathbb Z}
\def\C{\mathbb C}
\def\R{\mathbb R}

\usepackage{amsmath,amssymb}
\def\beq{\begin{equation}}
\def\eeq{\end{equation}}
\def\beqa{\begin{eqnarray}}
\def\eeqa{\end{eqnarray}}
\def\barray{\begin{array}}
\def\earray{\end{array}}

\def\sc{ \scriptscriptstyle \underline }

\def\sn{\mathrm{sn}}
\def\cn{\mathrm{cn}}
\def\dn{\mathrm{dn}}
\def\ns{\mathrm{ns}}
\def\nc{\mathrm{nc}}
\def\nd{\mathrm{nd}}
\def\sc{\mathrm{sc}}

\begin{document}

\title{
{\bf Exotic supersymmetry of the kink-antikink crystal, and
the infinite period limit}}

\author{
{\bf Mikhail S. Plyushchay${}^{a,b}$, Adri\'an
Arancibia${}^a$ and Luis-Miguel Nieto$^{b}$}
\\
[4pt] {\small \textit{${}^{a}$ Departamento de F\'{\i}sica,
Universidad de Santiago de Chile, Casilla 307, Santiago 2,
Chile  }}\\
{\small \textit{${}^{b}$ Departamento de F\'{\i}sica
Te\'orica, At\'omica y \'Optica, Universidad de Valladolid,
47071,
Valladolid, Spain}}\\
}

\date{}

\maketitle

\begin{abstract}
Some time ago, Thies et al. showed  that the Gross-Neveu model with
a bare mass term possesses a kink-antikink crystalline phase.
Corresponding self-consistent solutions, known earlier in  polymer
physics, are described by  a self-isospectral pair of one-gap
periodic Lam\'e potentials with a Darboux displacement depending on
the bare mass. We study an unusual supersymmetry of such a second
order Lam\'e system, and show that the associated first order
Bogoliubov-de Gennes Hamiltonian possesses the own nonlinear
supersymmetry. The Witten index is ascertained to be zero for  both
of the related exotic supersymmetric structures, each of which
admits several alternatives for the choice of a grading operator. A
restoration of the discrete chiral symmetry at zero value of the
bare mass, when the kink-antikink crystalline condensate transforms
into the kink crystal, is shown to be accompanied by structural
changes in both of the supersymmetries. We find that the infinite
period limit may or may not change the index. We also explain the
origin of the Darboux dressing phenomenon recently observed in a
non-periodic self-isospectral one-gap P\"oschl-Teller system, which
describes the  Dashen, Hasslacher and Neveu kink-antikink baryons.
\end{abstract}

\vskip.5cm\noindent

\section{Introduction}\label{introd}

The Gross-Neveu (GN) model \cite{GrNe,DaHN,NePa} is a remarkable
(1+1)-dimensional theory of self-interacting fermions that has no
gauge fields or gauge symmetries, but exhibits some important
features of quantum chromodynamics, namely, asymptotic freedom,
dynamical mass generation, and chiral symmetry breaking
\cite{Thies4}.  It has been widely studied over the years and the
richness of its properties is still astonishing. Some time ago,
Thies et al. showed that at finite density, the ground state of the
model with a discrete chiral symmetry is a kink crystal
\cite{Thies1}, while the kink-antikink crystalline phase was found
in the GN model with a bare mass term \cite{Thies2}. Then, Dunne and
Basar derived a new self-consistent inhomogeneous condensate, the
twisted kink crystal in the GN model with continuous chiral symmetry
\cite{BasDun1,BasDun2}. On the other hand, the relation of the GN
model with the sinh-Gordon equation and classical string solutions
in AdS${}_3$ has been observed recently \cite{Klotz,BasDun3}.

These two classes of the results seem to be different, but both are
rooted in the integrability features of the GN model, and may be
related to the Bogoliubov-de Gennes (BdG) equations incorporated
implicitly in its structure. It is because of these properties that
the model finds many applications in diverse areas of physics.
Particularly, the model  has provided very fruitful links between
particle and condensed matter physics, see  \cite{JackiwS,CamBish}
and \cite{Thies3}.

The origin of the model itself may also be somewhat related to the
BdG equations. We briefly discuss these equations to formulate  the
aim of the present paper. \vskip0.1cm

The BdG equations \cite{Bg} in the Andreev approximation \cite{Andr}
is a set of two coupled linear differential equations, which can be
presented in a form of a stationary Dirac-type matrix equation,
\begin{equation}\label{BdGfirst}
    \hat{G}_1\psi=\omega\psi,\qquad
    \hat{G}_1=a\sigma_1\frac{1}{i}\frac{d}{dx}-\sigma_2\Delta(x)\,.
\end{equation}
The scalar field $\Delta(x)$  is determined  via a self-consistency
condition, which often referred to as a gap equation. Equation
(\ref{BdGfirst}) arose in the theory of superconductivity by
linearizing the \emph{non-relativistic} energy dispersion (in
absence of magnetic field), or, equivalently, by neglecting  second
derivatives of the Bogoliubov amplitudes, see \cite{BarSKu}. A
constant $a$ is proportional there to the Fermi momentum $\hbar
k_F$.  In what follows we put $a=1$ and $\hbar=1$.

The Lagrangian of the GN model of the $N$ species of
self-interacting fermions is
\begin{equation}\label{LagGN}
\mathcal{L}_{GN}=\bar{\psi}(i\gamma^\mu\partial_\mu-m_0)\psi
+\frac{1}{2}g^2(\bar{\psi}\psi)^2\,,
\end{equation}
where $g^2$ is a coupling constant, the summation in the flavor
index is suppressed, and a bare mass term $\sim m_0$, which breaks
explicitly the discrete chiral symmetry
$\psi\rightarrow\gamma_5\psi$ of the massless model, is
included~\footnote{The investigation of model (\ref{LagGN}) is
motivated in \cite{Thies2} by a massive nature of quarks; there, the
't Hooft limit $N\rightarrow\infty$, $Ng^2=const$, is considered.}.
It is the two-dimensional version of the Nambu-Jona-Lasinio model
\cite{NJL} [with continuous chiral symmetry reduced to the discrete
one]. The latter is based on an analogy with superconductivity, and
was introduced as a model of symmetry breaking in particle physics.
There are two equivalent methods to seek for solutions for the GN
model. One of them is the Hartree-Fock approach, in which
self-consistent solutions to the Dirac equation
$(i\gamma^\mu\partial_\mu-\mathcal{S})\psi=0$ are looked for, with
spinor and scalar fields subject to a constraint of the form
$(\mathcal{S}(x)-m_0)=-Ng^2\langle\bar{\psi}\psi\rangle$, see
\cite{Thies4,Thies1,PTDol}. For static solutions, under appropriate
choice of the gamma matrices, the Dirac equation takes a form of the
BdG matrix equation (\ref{BdGfirst}), with $\hat{G}_1$ as a single
particle fermionic Hamiltonian. The condensate field
$\mathcal{S}(x)$ is identified with a gap function $\Delta(x)$,
while the constraint corresponds to the above mentioned gap
equation. Another approach to seek solutions for the GN model, in
which the BdG equations also play a key role, is via a functional
gap equation \cite{FeiZee,Dun4}. There, the condensate field is
given by stationary points of effective action, and  a connection of
the GN model with integrable hierarchies can be revealed, see
\cite{BasDun1,BasDun2,Dun4,CDP}. In light of this, the relation of
the GN model to the sinh-Gordon equation does not seem to be so
surprising as the BdG equations arise (in a slightly modified form)
as an important ingredient in solving the sine-Gordon equation, see
\cite{AKNS2,ChenHu}.

We now return  to the BdG matrix system (\ref{BdGfirst}). By
squaring, the equations decouple,
\begin{equation}\label{HSchBdG}
    \hat{H}\psi=E\psi,\qquad
    E=\omega^2,\qquad
    \hat{H}=-\frac{d^2}{dx^2}+\Delta^2
    -\sigma_3\Delta'\,.
\end{equation}
{}From the viewpoint of the second order system
$\hat{H}=\hat{G}_1^2$, the first order matrix operator $\hat{G}_1$
is a nontrivial integral of motion, $[\hat{H},\hat{G}_1]=0$. Having
also an integral $\sigma_3$, $[\hat{H},\sigma_3]=0$, which
anti-commutes with $\hat{G}_1$, we obtain a pattern of
supersymmetric quantum mechanics with $\sigma_3$ identified as a
grading operator. Though a system of the first and second order
equations (\ref{BdGfirst}) and (\ref{HSchBdG}) was exploited in
investigations on superconductivity, its superalgebraic structure,
which also includes the second supercharge
$\hat{G}_2=i\sigma_3\hat{G}_1$, seems to have gone unnoticed before
the theoretical discovery of supersymmetry in particle physics.
Supersymmetric quantum mechanics was then developed  by Witten as a
toy model for studying the supersymmetry breaking in quantum field
theories \cite{Wit}. Later, the relation of supersymmetric quantum
mechanics with Darboux transformations was noticed \cite{MatSal},
and found many applications \cite{SusyQm}. \vskip0.05cm

Braden and Macfarlane \cite{BraMac}, and, in a broader context,
Dunne and Feinberg \cite{DunFei} observed that the Darboux
transformed, supersymmetric partner of the one-gap periodic Lam\'e
system \cite{WW} with a zero energy ground state is described by the
same potential but translated for a half-period. The superpartner,
therefore, also has a zero ground state. Such a system is described
by unbroken supersymmetry, in which, however, the Witten index takes
zero value. For a class of superpesymmetric systems with
super-partner potentials of the same form a term
\emph{self-isospectrality} was coined by Dunne and Feinberg
\cite{DunFei}. The supersymmetric Lam\'e system considered in
\cite{BraMac,DunFei} corresponds to the kink crystalline phase
discussed in \cite{Thies1}, which describes a \emph{periodic}
generalization of the Callan-Coleman-Gross-Zee (CCGZ) kink
configurantions
 of the GN model, see \cite{DaHN,PTDol,Gross} and
\cite{BarSKu}. It was known earlier as a self-consistent
solution to the GN model in the context of  condensed
matter physics \cite{SaxBish}, see also
\cite{Braz,Horo,MachNak}.

The Lam\'e system, like non-periodic reflectionless solutions of the
GN model, belongs to a special class of the \emph{finite-gap}
systems \cite{MatSal,finegap}~\footnote{There is also a relation of
one-gap Lam\'e equation with the sine-Gordon equation, see
\cite{Suther}.}. Some time ago, it was found that such systems in an
unextended case (i.~e. when a second order Hamiltonian has a single
component), are characterized by a hidden, peculiar nonlinear
supersymmetry \cite{CP1,CP2}. It is associated with a corresponding
Lax operator (integral), and the grading is provided there by a
reflection operator. As a consequence, supersymmetric structure of
an extended system [with a matrix Hamiltonian of the form
(\ref{HSchBdG})] turns out to be  much richer than that associated
with only the first order supercharges $\hat{G}_a$, $a=1,2$, and
integral $\sigma_3$, see \cite{Tri}. It has also been shown recently
\cite{PRDPT} that the self-isospectral P\"oschl-Teller system (PT),
which describes the Dashen-Hasslacher-Neveu (DHN)  kink-antikink
baryons \cite{DaHN}, is characterized by a very unusual nonlinear
supersymmetric structure that admits six more alternatives for the
grading operator in addition to the usual choice of $\sigma_3$. All
the local and non-local supersymmetry generators turn out to be the
Darboux-dressed integrals of a free non-relativistic particle.
Moreover, it was shown there that the associated BdG system, with
the matrix operator (\ref{BdGfirst}) identified as a first order
(Dirac) Hamiltonian, possesses its own, nontrivial nonlinear
supersymmetry. \vskip0.05cm

In the present paper we investigate the exotic supersymmetric
structure of the kink-antikink crystal of \cite{Thies2,SaxBish},
which is a self-consistent solution of the GN model (\ref{LagGN})
with a real gap function $\Delta(x;\tau)$. Parameter $\tau$ is
related to $m_0$ and controls a central gap in the spectrum of the
first order BdG Hamiltonian operator (\ref{BdGfirst}).
Simultaneously, it defines a mutual displacement, $2\tau$, of
superpartner Lam\'e potentials in correspondence with the structure
of the second order Schr\"odinger operator (\ref{HSchBdG}). One more
parameter, not shown explicitly here, defines a period of the
crystal.  A quarter-period value of $\tau$ corresponds to the kink
crystal solution of \cite{Thies1} for the model (\ref{LagGN}) with
$m_0=0$, which was considered in \cite{BraMac,DunFei}.  We also
study different forms of the infinite-period limit applied to the
supersymmetric structure. \emph{A priori} the picture of such a
limit has to be rather involved\,: the Darboux dressing relates the
non-periodic kink-antikink system to a free particle, while the
Darboux transformations in the periodic case are expected to be just
self-isospectral displacements, see \cite{SaxBish,Tri,FerNN,Sams}.
\vskip0.05cm

The outline of the paper is as follows. In the next section, we
discuss the main properties of the one-gap Lam\'e system. In section
3 we construct its self-isospectral extension by employing certain
eigenfunctions of the Lam\'e Hamiltonian. We investigate the action
of the first order Darboux displacement generators, and discuss the
spectral peculiarities of the obtained supersymmetric system.
Section 4 is devoted to the study of the properties of a
superpotential (gap function) that is an elliptic function both in a
variable and a shift parameter. These properties are employed in
section 5, where we construct the second order intertwining
operators, identify further local matrix integrals of motion, and
compute a corresponding nonlinear superalgebra. In section 6 we show
that the system possesses six more, nonlocal integrals of motion,
each of which may be chosen as a $\Z_2$ grading operator instead of
a usual integral $\sigma_3$ of the supersymmetric quantum mechanics.
We discuss alternative forms of the superalgebra associated with
these additional integrals and their action on the physical states
of the system. In section 7, we investigate a peculiar nonlinear
supersymmetry of the associated first order BdG system. Section 8 is
devoted to the infinite period limit of the both, second and first
order supersymmetric systems. In section 9 we clarify the origin of
the Darboux dressing phenomenon that takes place in the non-periodic
self-isospectral PT system, that was revealed in \cite{PRDPT}. In
section  10 we discuss the obtained results. To provide a
self-contained presentation, the necessary properties of Jacobi
elliptic functions and of some related non-elliptic functions are
summarized in the two appendices.


\section{One-gap Lam\'e equation}

In this section we discuss the properties of the Lam\'e system which
is necessary for further constructions and analysis.

\begin{figure}[h!]\begin{center}
\includegraphics[scale=1.2]{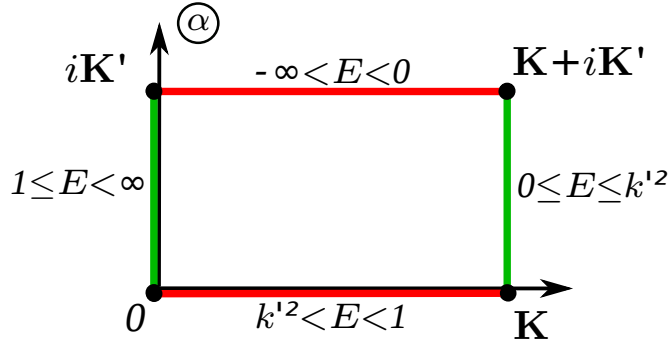}
\caption{ The sides of the rectangle  are mapped by (\ref{Ealpha})
onto the indicated energy intervals. The vertical (horizontal) sides
shown in green (red) correspond to the two allowed (forbidden)
bands. Vertices $\alpha={\rm {\bf K}}+i{\rm {\bf K}}'$, ${\rm {\bf
K}}$ and $0$ are mapped, respectively, into the edges $E=0$,
$k'{}^2$, and $1$ of the valence, $0\leq E\leq k'^2$, and
conduction, $1\leq E<\infty$, bands, which are described by
periodic, $\dn\,x$ ($E=0$), and antiperiodic, $\cn\,x$ ($E=k'^2$)
and $\sn\,x$ ($E=1$), functions. Vertex $i{\rm {\bf K}}'$ as a limit
point on a horizontal (vertical) side corresponds to $E=-\infty$
($E=+\infty$). }\label{fig1}
\end{center}
\end{figure}

Consider the simplest (and unique) \emph{one-gap} periodic second
order system described by the Lam\'e Hamiltonian
\begin{equation}\label{HLame}
    H=-\frac{d^2}{dx^2}+2k^2\sn^2x-k^2\,.
\end{equation}
An additive  constant term is chosen here such that a minimal energy
value (the lower edge of the valence band, see below) is zero.
Potential $V(x)=2k^2\sn^2x-k^2$ is a periodic function with a real
period $2{\rm {\bf K}}$ (and a pure imaginary period $2i{\rm {\bf
K}}'$)~\footnote{See Appendices A and B for notations and properties
we use for Jacobi elliptic and related functions.}. The general
solution of the equation
\begin{equation}\label{SchEq}
    H\Psi(x)=E\Psi(x)
\end{equation}
is given by \cite{WW}
\begin{equation}\label{Psipm}
    \Psi_\pm^{\alpha}(x)=\frac{{\rm H}(x\pm \alpha)}{\Theta(x)}
    \exp\left[
    \mp x {\rm Z}(\alpha)\right]\,.
\end{equation}
Here ${\rm H}$, $\Theta$ and ${\rm Z}$ are  Jacobi's Eta, Theta and
Zeta functions,  and the eigenvalue $E=E(\alpha)$ is defined by the
relation
\begin{equation}\label{Ealpha}
    E(\alpha)=\dn^2\alpha\,.
\end{equation}
The Hamiltonian (\ref{HLame}) is Hermitian, and we treat
(\ref{SchEq}) as the stationary Schr\"odinger equation on a real
line. We are interested in the values of the parameter $\alpha$,
which give real $E$. $\dn^2 \alpha$ is an elliptic function with
periods $2{\rm {\bf K}}$ and $2i{\rm {\bf K}}'$, and its period
parallelogram in a complex plane is a rectangle with vertices in
$0$, $2{\rm {\bf K}}$, $2{\rm {\bf K}}+ 2i{\rm {\bf K}}'$ and
$2i{\rm {\bf K}}'$. We then look for those $\alpha$ in the period
parallelogram for which $\dn\,\alpha$ takes real or pure imaginary
values. They can be taken, for instance, on the border of the
rectangle shown on Fig. \ref{fig1}.
We have, particularly,
\begin{equation}\label{E1band1}
    E({\rm {\bf K}}+i\beta)=
    k'{}^2{\cn^2(\beta|k')}{\nd^2(\beta|k')}\,,
    \qquad
    0\leq \beta\leq {\rm {\bf K}}'\,,\quad
    k'{}^2\geq E({\rm {\bf K}}+i\beta)\geq 0\,,
\end{equation}
\begin{equation}\label{E1band2}
    E(i\beta)={\dn^2(\beta|k')}{\nc^2(\beta|k')}=
    k'{}^2+{k^2}{\nc^2(\beta|k')}\,,
    \qquad
    0\leq\beta<{\rm {\bf K}}'\,,\quad
    1\leq E(i\beta)<\infty\,.
\end{equation}
For (\ref{E1band1}) and (\ref{E1band2}), eigenfunctions in
(\ref{SchEq}) are bounded on a real line, that corresponds
to the two allowed (valence and conduction) bands in the
spectrum. In contrast, for $\alpha=\beta$ and
$\alpha=\beta+i{\rm {\bf K}}'$, $\beta\in (0,{\rm {\bf
K}})$, a real part of ${\rm Z}(\alpha)$ is nonzero, and
eigenfunctions (\ref{Psipm}) are not bounded for
$|x|\rightarrow \infty$. This corresponds to the two
forbidden zones, $-\infty<E<0$ and $k'{}^2<E<1$.

Differentiation of (\ref{E1band1}) and (\ref{E1band2}) in
$\beta$ gives a relation
\begin{equation}\label{dEdb}
    \frac{dE}{d\beta}=2\eta(E)\sqrt{P(E)}\,,\qquad
    P(E)=E(E-k'{}^2)(E-1)\,.
\end{equation}
The third order polynomial $P(E)$ takes positive values
inside the allowed bands, and turns into zero at their
edges.  $\eta(E)$ takes values $-1$ and $+1$ in the valence
and conduction bands, respectively.

Inside the two allowed bands, (\ref{Psipm}) are
quasi-periodic Bloch wave functions,
\begin{equation}\label{quasimom}
   \Psi^\alpha_\pm(x+2{\rm {\bf K}})=
    e^{\mp i2{\rm {\bf K}}\kappa(E)}\Psi^\alpha_\pm(x)\,,\qquad
    \kappa(E)=\frac{\pi}{2{\rm {\bf K}}}-i{\rm
    Z}(\alpha)\,,
\end{equation}
where a first term in quasimomentum (crystal momentum)
$\kappa(E)$ originates from the imparity of ${\rm H}$
function.  In the valence, (\ref{E1band1}),  and
conduction, (\ref{E1band2}), bands its values are given by
\begin{eqnarray}\label{kappa1}
    &\kappa(E({\rm {\bf K}}+i\beta))=
    \frac{\pi}{2{\rm {\bf K}}}
    -\left[{\rm Z}(\beta|k')+
    \frac{\pi}{2{\rm {\bf K}}{\rm {\bf K}}'}\beta -
    k'{}^2
    \cn(\beta|k')\sn(\beta|k')\nd(\beta|k')\right]\,,&\\
    &\kappa(E(i\beta))=\frac{\pi}{2{\rm {\bf K}}}
    -\left[{\rm Z}(\beta|k')+
    \frac{\pi}{2{\rm {\bf K}}{\rm {\bf K}}'}\beta -
    \dn(\beta|k')\sn(\beta|k')\nc(\beta|k')\right]\,.&
    \label{kappa2}
\end{eqnarray}
With the help of (\ref{Ealpha}) and (\ref{dEdb}) one finds
a differential dispersion relation
\begin{equation}\label{dkdE}
    \frac{d\kappa}{dE}=
    \eta(E)\,\frac
    { E-({\rm {\bf E}}/{\rm {\bf
    K}})}{2\sqrt{P(E)}}\,,
\end{equation}
where ${\rm \bf{E}}$ is a complete elliptic integral of the
second kind, see (\ref{EK}). Taking into account a relation
$k'{}^2<\frac{{\rm {\bf E}}}{{\rm {\bf K}}}<1$, see
Appendix B, one finds that within the both allowed bands
quasimomentum is increasing function of energy. It takes
values $0$ and $\pi/2{\rm {\bf K}}$ at the edges $E=0$ and
$E=k'{}^2$  of the valence band, where Bloch-Floquet
functions reduce to the periodic, $\dn\, x$, and
antiperiodic, $\cn\, x$, functions in the real period
$2{\rm {\bf K}}$ of the system. Within the conduction band,
quasi-momentum increases from ${\pi}/{2{\rm {\bf K}}}$ to
$+\infty$. At the lower edge $E=1$, two functions
(\ref{Psipm}) reduce to the antiperiodic function $\sn\,
x$. At all three edges of the allowed bands, derivative of
quasimomentum in energy is $+\infty$. For large values of
energy, $E\rightarrow +\infty$, we find that
$\kappa(E)\approx \sqrt{E}$, i.e. Bloch functions
(\ref{Psipm}) behave as the plane waves,
$\Psi_{\pm}^\alpha(x+2{\rm {\bf K}})\approx e^{\mp i2{\bf
{\rm K}}\sqrt{E}}\Psi_{\pm}^\alpha(x)$.

Second, linear independent solutions at the edges of the
allowed bands $E_i=0,\,k'{}^2,\,1$ are
$\Psi_i(x)=\psi_i(x)\mathcal{I}_i$, where
$\mathcal{I}_i=\int dx/\psi_i^2(x)$, and $\psi_i=\dn\, x$,
$\cn\,x$, $\sn\, x$, $i=1,2,3$. The integrals  are
expressed in terms of non-periodic incomplete elliptic
integral of the second kind (\ref{E(u)}),
$\mathcal{I}_1=\frac{1}{k'{}^2}{\rm E}(x+{\rm {\bf K}})$,
$\mathcal{I}_2=x-\frac{1}{k'{}^2}{\rm E}(x+{\rm {\bf
K}}+i{\rm {\bf K}}')$, $\mathcal{I}_3=x-{\rm E}(x+i{\rm
{\bf K}}')$. $\Psi_i(x)$ are not bounded on $\R$ and
correspond to non-physical states. These non-physical
solutions follow also from general solutions (\ref{Psipm}).
For instance, $\Psi_3(x)$ may be obtained as a limit of
$(\Psi^\alpha_+(x)-\Psi^\alpha_-(x))/\alpha$ as
$\alpha\rightarrow 0$. Eq. (\ref{Psipm}) provides a
complete set of solutions for (\ref{SchEq}) as the second
order differential equation. Notice also that Bloch states
(\ref{Psipm}) within the allowed bands are related under
complex conjugation as
$(\Psi^\alpha_+(x))^*=\eta\Psi^\alpha_-(x)$, where $\eta$
is the same as in (\ref{dEdb}).

In conclusion of this section we note that the function
$P(E)$ in Eqs. (\ref{dEdb}), (\ref{dkdE}) is a
\emph{spectral polynomial}. It will play a fundamental role
in a nonlinear supersymmetry we will discuss below.

\section{Self-isospectral Lam\'e system}

Consider the lower in energy $E$ forbidden band by extending it with
the edge value $E=0$ of the valence band. We introduce a notation
$-2\tau+i{\rm {\bf K}}'$ for the parameter $\alpha$ that corresponds
to the extended interval $-\infty<E\leq 0$. By taking into account
relations $\dn\,(-u)=\dn\,(u+2{\rm {\bf K}})=-\dn\,(u+2i{\rm {\bf
K}}')=\dn\,u$, it will be convenient do not restrict the values of
$\tau$ to the interval $[-{\rm {\bf K}}/2,0)$, but assume that
$\tau\in\R$, while keeping in mind that $E\rightarrow -\infty$ for
$\tau\rightarrow n{\rm {\bf K}}$, $n\in\Z$. After a shift of the
argument $x\rightarrow x+\tau$, the corresponding function
$\Psi_+^\alpha$ from (\ref{Psipm}) with $\alpha=-2\tau+i{\rm {\bf
K}}'$ takes, up to an inessential multiplicative constant, the form
\begin{equation}\label{psi+t}
    \frac{\Theta(x_-)}{\Theta(x_+)}
    \exp{\left[x\,{\rm z}(\tau)\right]}\equiv F(x;\tau)\,,
\end{equation}
where we have introduced the notations $x_+=x+\tau$, $x_-=x-\tau$,
\begin{equation}\label{c02t}
    {\rm z}(\tau)=-i\kappa(E(-2\tau+i{\rm {\bf
    K}}'))=
    \varsigma(\tau)+{\rm Z}(2\tau)=\frac{1}{2}
    \frac{d}{d\tau}\ln(\Theta(2\tau)\,\sn\,2\tau)\,,
\end{equation}
\begin{equation}\label{c02t+}
    \varsigma(\tau)=\frac{1}{2}\frac{d}{d\tau}\ln\sn\,2\tau=\ns\,2\tau\,\cn\,
2\tau\,\dn\,2\tau\,.
\end{equation}
$F(x;\tau)$ is a quasi-periodic in $x$ and periodic in the $\tau$
function, $
    F(x+2{\rm {\bf K}};\tau)=\exp{\big(2{\rm {\bf K}}
    {\rm z}(\tau)\big)}F(x;\tau),$
$
    F(x;\tau+2{\rm {\bf K}})=F(x;\tau).
$ It is a regular function of $\tau$, save for $\tau=n{\rm {\bf
K}}$, $n\in \Z$, [which correspond to the poles $\alpha=2n{\rm {\bf
K}}+i{\rm {\bf K}}'$ of $\dn\,\alpha$ in (\ref{Ealpha})], where
$F(x;\tau)$ with $x\neq 0$ undergoes infinite jumps from $0$ to
$+\infty$. Since ${\rm z}({\rm {\bf K}}/2)=0$, function
(\ref{psi+t}) reduces at $\tau={\rm {\bf K}}/2$ (up to an
inessential multiplicative constant) to a periodic in the $x$
function $\dn\,(x+\frac{1}{2}{\rm {\bf K}})$  which describes a
physical state with energy $E=0$ at the lower edge of the valence
band of the system $H(x+\frac{1}{2}{\rm {\bf K}})$. $F(x;\tau)$ is a
nodeless function that obeys the relations $F(x;-\tau)=
F(-x;\tau)=1/F(x;\tau)$ and
\begin{equation}\label{Hve2t}
    \left[H(x_+)+\varepsilon(\tau)
    \right]F(x;\tau)=0\,,\quad
    {\rm where}\quad
    \varepsilon(\tau)=-E(-2\tau+i{\rm {\bf K}}')=
    \cn^2\, 2\tau\,\ns^2\, 2\tau\,.
\end{equation}

Define a first order differential operator
\begin{equation}\label{Delsuper}
    \mathcal{D}(x;\tau)=
    F(x;\tau)\,\frac{d}{dx}\,\frac{1}{F(x;\tau)}=
    \frac{d}{dx}-\Delta(x;\tau)\,,\qquad
    \mathcal{D}^\dagger(x;\tau)=-\mathcal{D}(x;-\tau)\,,
\end{equation}
where
\begin{equation}\label{Dsupergap}
    \Delta(x;\tau)
    =\frac{F'(x;\tau)}{F(x;\tau)}\,.
\end{equation}
Operator (\ref{Delsuper}) annihilates function
(\ref{psi+t}), $\mathcal{D}(x;\tau)F(x;\tau)=0$, and we
find that
\begin{equation}\label{Hx+t}
    \mathcal{D}^\dagger(x;\tau)\mathcal{D}(x;\tau)=
H(x_+)+\varepsilon(\tau)\,,\qquad
    \mathcal{D}(x;\tau)\mathcal{D}^\dagger(x;\tau)=
      H(x_-)+\varepsilon(\tau)\,.
\end{equation}
By virtue of $\varepsilon(\frac{1}{2}{\rm {\bf
K}})=0$, a non-shifted Lam\'e Hamiltonian operator
(\ref{HLame}) factorizes then as
$ H(x)=\mathcal{D}\left(x+\frac{1}{2}{\rm {\bf
K}};\frac{1}{2}{\rm {\bf K}}\right)
    \mathcal{D}^\dagger\left(x+\frac{1}{2}{\rm {\bf K}};
\frac{1}{2}{\rm {\bf K}}\right)$.  The alternative product
produces a shifted in the half-period ${\rm {\bf K}}$
system, $ H(x+{\rm {\bf
K}})=\mathcal{D}^\dagger\left(x+\frac{{\rm {\bf
K}}}{2};\frac{1}{2}{\rm {\bf K}}\right)
    \mathcal{D}\left(x+\frac{1}{2}{\rm {\bf K}};
\frac{1}{2}{\rm {\bf K}}\right)\,. $ It is this
factorization of a pair of  Lam\'e Hamiltonians $H(x)$ and
$ H(x+{\rm {\bf K}})$  that underlies a usual
supersymmetric structure studied in \cite{DunFei} in the
light of a phenomenon of self-isospectrality.

Notice that while $F(x_-;\tau)$ is, up to a multiplicative
constant, a non-physical eigenfunction
$\Psi^{-2\tau+iK'}_+(x)$ of $H(x)$ of energy
$-\varepsilon(\tau)$, function $F(x_+;-\tau)=1/F(x_+;\tau)$
coincides, up to a multiplicative constant, with another
eigenfunction $\Psi^{-2\tau+iK'}_-(x)$ of $H(x)$ with the
same eigenvalue.

\begin{figure}[h!]\begin{center}
\includegraphics[scale=0.8]{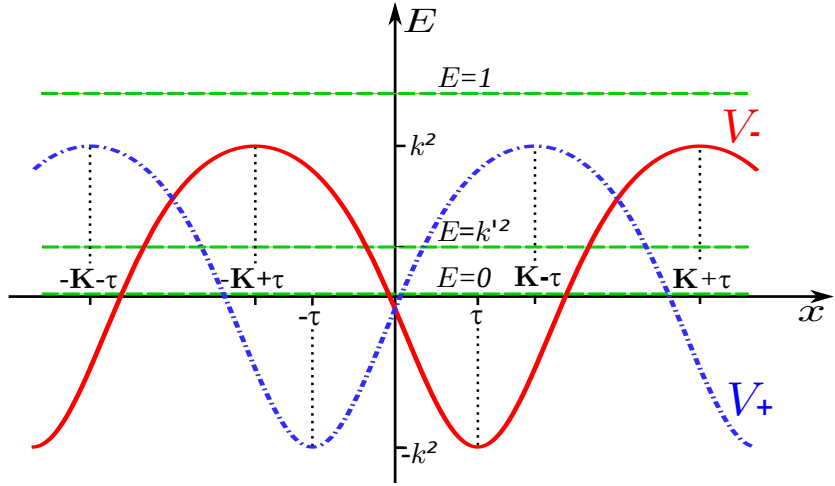}
\caption{ The self-isospectral potentials $V_\pm=2k^2\sn(x_\pm)-k^2$
are shown together with the edges of the valence ($0\leq E\leq
k'{}^2$) and conduction ($1\leq E<\infty$) bands.  $V_\pm$ have
maxima at $x=\mp\tau+(2n+1){\rm {\bf K}}$ and minima at
$x=\mp\tau+2n{\rm {\bf K}}$. Here $k^2=0.75$, ${\rm {\bf K}}=2.16$,
and $\tau=0.8$. }\label{fig2}
\end{center}
\end{figure}

According to (\ref{Hx+t}), the mutually shifted
Hamiltonians $H(x+\tau)$ and $H(x-\tau)$ form a
supersymmetric, self-isospectral periodic one-gap Lam\'e
system
\begin{equation}\label{Hcal}
    \mathcal{H}={\rm diag}\, (H(x_+), H(x_-)),
\end{equation}
see Fig. \ref{fig2}, for which $\Delta(x;\tau)$ plays the role of
the superpotential that obeys the Ricatti equations
\begin{equation}\label{Ricatti}
    \Delta^2(x;\tau)\pm \Delta'(x;\tau) =2k^2
\sn^2(x \pm\tau)-k^2+\varepsilon(\tau)\,.
\end{equation}
Indeed, from factorizations (\ref{Hx+t}) it follows that
the $\mathcal{D}(x;\tau)$ and $\mathcal{D}^\dagger(x;\tau)$
intertwine the Hamiltonians $H(x_+)$ and $H(x_-)$,
\begin{equation}\label{DHinter}
    \mathcal{D}(x;\tau)H(x_+)=
    H(x_-)\mathcal{D}(x;\tau)\,,\qquad
    \mathcal{D}^\dagger(x;\tau)H(x_-)=
    H(x_+)\mathcal{D}^\dagger(x;\tau)\,,
\end{equation}
and interchange the eigenstates of the superpartner
systems,
\begin{equation}\label{Dintersusy}
    \mathcal{D}(x;\tau)\Psi^\alpha_\pm(x_+)
    =\mathcal{F}^{\mathcal{D}}_\pm (\alpha,\tau)\,
    \Psi^\alpha_\pm(x_-)\,,\quad
    \mathcal{D}^\dagger(x;\tau)\Psi^\alpha_\pm(x_-)
    =-\mathcal{F}^{\mathcal{D}}_\pm (\alpha,-\tau)\,
    \Psi^\alpha_\pm(x_+)\,.
\end{equation}
The second relation in (\ref{Dintersusy}) follows from the
first one via a substitution $\tau\rightarrow -\tau$.
 A complex amplitude,
$\mathcal{F}^{\mathcal{D}}_\pm (\alpha,\tau)=e^{\pm
    i\varphi^{\mathcal{D}}(\alpha,\tau)}\mathcal{M}^{\mathcal{D}}(\alpha,\tau)$,
is given by
\begin{equation}\label{FD}
    \mathcal{F}^{\mathcal{D}}_\pm(\alpha,\tau)=
    -\exp
    \left[\mp 2i\left(\kappa(\alpha)
    -\frac{\pi}{2{\rm {\bf K}}} \right)\tau\right]\ns\,2\tau\,
    \frac{\Theta(2\tau\pm
    \alpha)\,\Theta(0)}{\Theta(2\tau)\Theta(\alpha)}\,.
\end{equation}
It satisfies  $
    \left(\mathcal{F}^{\mathcal{D}}_\pm(\alpha,\tau)\right)^*=
    \mathcal{F}^{\mathcal{D}}_\mp(\alpha,\tau)=
    -\mathcal{F}^{\mathcal{D}}_\pm(\alpha,-\tau).
$ Its modulus may be presented in a form
$\mathcal{M}^{\mathcal{D}}(\alpha,\tau)=
    \sqrt{E(\alpha)+
    \varepsilon(\tau)}\,,
$ where $E(\alpha)$ for the valence and conduction bands is
given by Eqs. (\ref{E1band1}) and (\ref{E1band2}). This
agrees with Eq. (\ref{Hx+t}).  Notice that the modulus is
an even in $\tau$ function,
$\mathcal{M}^{\mathcal{D}}(\alpha,\tau)=
\mathcal{M}^{\mathcal{D}}(\alpha,-\tau)$, which is nonzero
except for the lower edge states of the valence band
($E=0$) in the case $\tau=(\frac{1}{2}+n){\rm {\bf K}}$. A
phase is well defined  for $\mathcal{M}^{\mathcal{D}}\neq
0$, and satisfies a relation
\begin{equation}\label{varphiD}
    e^{i\varphi^{\mathcal{D}}(\alpha,-\tau)}=-e^{-i\varphi^{\mathcal{D}}\,
    (\alpha,\tau)}\,.
\end{equation}
It can be presented in a form
\begin{equation}\label{phaseexplicit}
    e^{i\varphi^{\mathcal{D}}(\alpha,\tau)}=
    -{\rm sign}\,(\ns\,2\tau)\exp
    \left[ -2i\left(\kappa(\alpha)
    -\frac{\pi}{2{\rm {\bf K}}} \right)\tau+i
    \varphi_\Theta(\alpha,\tau)\right],
\end{equation}
where ${\rm sign}\,(.)$ is a sign function, and
$\varphi_\Theta(\alpha,\tau)$ is a phase of
$\Theta(2\tau+\alpha)$,
$
    \varphi_\Theta(\alpha,\tau)=
    {\rm Im}\left( \int_0^{2\tau+\alpha}Z(u)\,du\right),
$ see Eq. (\ref{Theta}). Particularly, for the edge states
($i=1,2,3$), Eq. (\ref{FD}) gives $
    \mathcal{D}(x;\tau)\psi_i(x_+)=
    \mathcal{F}^{\mathcal{D}}_i(\tau)\psi_i(x_-)$,
$    \mathcal{D}^\dagger(x;\tau)\psi_i(x_-)=
    \mathcal{F}^{\mathcal{D}}_i(\tau)\psi_i(x_+)$,
where
\begin{equation}\label{psiiCi}
    \psi_i(x)=\dn\,x\,,\,\,\cn\,x\,,\,\,\sn\,x\,,\quad
    \mathcal{F}^{\mathcal{D}}_i(\tau)=
    -\cn\,2\,\tau\,\ns\,2\tau\,,\,\,-\dn\,2\,\tau\,\ns\,2\tau\,,\,\,
    -\ns\,2\tau\,,
\end{equation}
and so,
\begin{equation}\label{Medge}
    \mathcal{M}^{\mathcal{D}}_i(\tau)=\sqrt{\varepsilon(\tau)}\,,\,\,
    \sqrt{k'{}^2+\varepsilon(\tau)}\,,\,\,
    \sqrt{1+\varepsilon(\tau)}\,,
\end{equation}
and $ e^{i\varphi_i^\mathcal{D}(\tau)}=-{\rm
    sign}\,(\cn\,2\tau\,\ns\,2\tau)$,
    $-{\rm
    sign}\,(\ns\,2\tau),$
    $-{\rm
    sign}\,(\ns\,2\tau).$

As a consequence of intertwining relations (\ref{DHinter}),
first order matrix operators
\begin{equation}\label{Sdef}
    S_1=\left(%
    \begin{array}{cc}
  0 & \mathcal{D}^\dagger(x;\tau) \\
  \mathcal{D}(x;\tau) & 0 \\
    \end{array}%
    \right),\qquad
    S_2=i\sigma_3S_1\,,
\end{equation}
are the integrals of motion for system (\ref{Hcal}).
Integrals (\ref{Sdef})  correspond here (up to a unitary
transformation of sigma  matrices) to the first order
operators $\hat{G}_a$ in section \ref{introd}. Operator
$\Gamma=\sigma_3$ is a trivial integral for (\ref{Hcal}),
$[\Gamma,\mathcal{H}]=0$, that anticommutes with $S_a$,
$a=1,2$, $\{\Gamma,S_a\}=0$, and classifies them as
supercharges. Bosonic, $\mathcal{H}$, and fermionic, $S_a$,
operators satisfy then the $N=2$ supersymmetry algebra,
\begin{equation}\label{SSH}
    \{S_a,S_b\}=2\delta_{ab}(\mathcal{H}+\varepsilon(\tau))\,,\qquad
    [\mathcal{H},S_a]=0\,.
\end{equation}

In correspondence with (\ref{Dintersusy}) and
(\ref{varphiD}), the eigenstates of the supercharge $S_1$
are
\begin{equation}\label{S1eigen}
    S_1\Psi^\alpha_{\pm,S_1,\epsilon}=
    \epsilon\mathcal{M}^{\mathcal{D}}(\alpha,\tau)
    \Psi^\alpha_{\pm,S_1,\epsilon}\,,\quad
    \Psi^\alpha_{\pm,S_1,\epsilon}=
    \left(%
\begin{array}{c}
  \Psi^\alpha_\pm(x_+) \\
  \epsilon e^{\pm i\varphi^\mathcal{D}(\alpha,\tau)}
    \Psi^\alpha_\pm(x_-) \\
\end{array}%
\right),\quad \epsilon=\pm 1\,.
\end{equation}

Since $\varepsilon(\tau)>0$  for $\tau\neq (\frac{1}{2}+n){\rm {\bf
K}}$, $n\in\Z$,  the first-order supersymmetry
(\ref{SSH})~\footnote{ This refers to the order of the polynomial in
$\mathcal{H}$ that appears in the anticommutator of the
supercharges.} is dynamically broken in general case. It is unbroken
however for $\tau=(n+\frac{1}{2}){\rm {\bf K}}$ by virtue of
$\varepsilon((\frac{1}{2}+n){\rm {\bf K}})=0$. For these values of
the shift parameter, the supercharges $S_a$ annihilate the ground
states $\dn\,(x+(n+\frac{1}{2}){\rm {\bf K}})$ and
$\dn\,(x-(n+\frac{1}{2}){\rm {\bf K}})$ of the super-partner systems
$H(x+(n+\frac{1}{2}){\rm {\bf K}})$ and $H(x-(n+\frac{1}{2}){\rm
{\bf K}})$. Notice that with variation of the shift parameter
$\tau\neq n{\rm {\bf K}}$, which simultaneously governs the scale of
the supersymmetry breaking $\varepsilon(\tau)$, the spectrum of the
second order system (\ref{Hcal}) does not change. Each of its two
super-partners has the same spectrum as a non-shifted Lam\'e system
(\ref{HLame}) does. Therefore, each energy level inside the valence,
$0<E<k'{}^2$,  and conduction, $1<E<\infty$, bands is fourth-fold
degenerate in accordance with the existence of the two Bloch states,
$\Psi^\alpha_\pm(x_+)$ and $\Psi^\alpha_\pm(x_-)$, of the form
(\ref{Psipm}) for each subsystem, see Eq. (\ref{S1eigen}). We have a
two-fold degeneration at the edges $E=0$, $E=k'{}^2$ and $E=1$ of
the valence and conduction bands  in the spectrum of the
supersymmetric system $\mathcal{H}$. Bosonic, $\Psi^{(+)}$, and
fermionic, $\Psi^{(-)}$, states are defined as eigenstates of the
grading operator $\Gamma=\sigma_3$,
$\Gamma\Psi^{(\pm)}=\pm\Psi^{(\pm)}$, and have the general form
$\Psi^{(+)}=(\Psi(x_+),0)^T$ and $\Psi^{(-)}=(0,\Psi(x_-))^T$, where
$T$ means a transposition. In summary, we see that in both the
broken and unbroken cases, the Witten index, which characterizes the
difference between the number of bosonic and fermionic zero modes,
is the same and equals zero.

For $\tau\neq (\frac{1}{2}+n){\rm {\bf K}}$ (when
$\varepsilon(\tau)\neq 0$) supersymmetric relations
(\ref{SSH}) look differently from a usual form of
superalgebra in supersymmetric quantum mechanics. A simple
redefinition of the matrix Hamiltonian (\ref{Hcal}),
$\mathcal{H}\rightarrow
\tilde{\mathcal{H}}=\mathcal{H}+\varepsilon(\tau)$, will
correct the form of superalgebraic relations, but will not
change the conclusions on a broken (for $\tau\neq
(\frac{1}{2}+n){\rm {\bf K}}$) form of the supersymmetric
structure that we have analyzed.  We shall return to this
point in the discussion of the peculiar supersymmetry of
the first order Bogoliubov-de Gennes system in  section
\ref{susyBdGsec}. \vskip0.05cm

The described degeneracy of the energy levels in both, broken and
unbroken, cases is unusual for $N=2$ supersymmetry. We will show
that additional nontrivial integrals of motion may be associated
with this peculiarity of the self-isospectral supersymmetric system
(\ref{Hcal}). To identify such integrals, in the next section we
investigate the function $\Delta(x;\tau)$ in greater detail.

\section{Superpotential}

Being the logarithmic derivative of $F(x;\tau)$, see Eq.
(\ref{Dsupergap}), the superpotential $\Delta(x;\tau)$ may be
written with the help of (\ref{ZTheta}), (\ref{elThH}) in terms of
Jacobi's ${\rm Z}$, or $\Theta$ and ${\rm H}$ functions,
\begin{equation}\label{DelZeta}
    \Delta(x;\tau)={\rm z}(\tau)+{\rm Z}(x_-)
    -{\rm Z}(x_+)=\frac{1}{2}\frac{\partial}{\partial\tau}\ln\left(
    \frac{{\rm H}(2\tau)}{\Theta^2(x_-)\Theta^2(x_+)}
    \right)\,.
\end{equation}
The addition formula (\ref{Zadd}) for the ${\rm Z}$ function gives
another, equivalent representation
\begin{equation}\label{Del1}
    \Delta(x;\tau)=\varsigma(\tau)
    +k^2\sn\, 2\tau\,\sn\,(x_-)\,\sn\,(x_+)\,.
\end{equation}
Functions ${\rm z}(\tau)$ and $\varsigma(\tau)$ are defined
in (\ref{c02t}), (\ref{c02t+}).  Yet another useful
representation for the superpotential may be derived from
(\ref{Del1}),
\begin{equation}\label{Delta3}
    \Delta(x;\tau)=\frac{\sn\,x_-\cn\,x_-\dn\,x_-
    +
    \sn\,x_+\cn\,x_+\dn\,x_+}{
    \sn^2 x_+-\sn^2x_-}\,.
\end{equation}

 Having in mind relations (\ref{DHinter}),
(\ref{Hx+t}) and (\ref{Ricatti}), in what follows we treat $x$ as a
variable and $\tau$ as a shift parameter. $\Delta(x;\tau)$ is an
elliptic function in both its arguments with the same periods $2{\rm
{\bf K}}$ and $2i{\rm {\bf K}}'$. It is an \emph{even} in $x$ and
\emph{odd} in the $\tau$ function with respect to the points $0,K$
(modulo periods), $\Delta(-x;\tau)=\Delta(x;\tau)$, $\Delta({\rm
{\bf K}}-x;\tau)=\Delta({\rm {\bf K}}+x;\tau)$,
$\Delta(x;-\tau)=-\Delta(x;\tau)$, $\Delta(x;{\rm {\bf
K}}-\tau)=-\Delta(x;{\rm {\bf K}}+\tau)$. It also obeys a relation
$\Delta(x+{\rm {\bf K}};\tau+{\rm {\bf K}})=\Delta(x-{\rm {\bf
K}};\tau+{\rm {\bf K}})=\Delta(x;\tau)$. In $\tau=0,{\rm {\bf K}}$
the function undergoes infinite jumps.

Being the elliptic function in $x$, $\Delta(x;\tau)$ obeys a
nonlinear differential equation
\begin{equation}\label{DelEq1}
    {\Delta'}^2=\Delta^4+2\delta_2(\tau)\Delta^2
    +\delta_1(\tau)\Delta+\delta_0(\tau)\,,
\end{equation}
where $\delta_2(\tau)=1+k^2-3\ns^2 \,2\tau$,
$\delta_1(\tau)=8\ns^3\, 2\tau\,\cn\,2\tau\,\dn\,2\tau$, and $
\delta_0(\tau)=-3\ns^4\, 2\tau + 2(1+k^2)\ns^2\, 2\tau+k'{}^4$. As a
consequence of (\ref{DelEq1}), it also satisfies the nonlinear
higher order differential equations
\begin{equation}\label{DelEq23}
    \Delta''=2\Delta^3+2\delta_2(\tau)\Delta+\frac{1}{2}
    \delta_1(\tau)\,,\qquad
    \Delta'''=2\Delta'\left(3\Delta^2+\delta_2(\tau)\right)\,.
\end{equation}

Making use of (\ref{DelZeta}), one finds the relation
\begin{equation}\label{Deladd}
    \Delta(x+\tau+\lambda;\lambda)-\Delta(x+\lambda;\tau+\lambda)+
    \Delta(x;\tau)=g(\tau,\lambda)\,.
\end{equation}
Function
$g(\tau,\lambda)=\varsigma(\tau)+\varsigma(\lambda)-
    \varsigma(\tau+\lambda)+
    k^2\sn\, 2\tau\,\sn\,2\lambda\,\sn\,2(\tau+\lambda)$
has symmetry properties
$g(\tau,\lambda)=g(\lambda,\tau)=g(\tau,-\lambda-\tau)=
    -g(-\tau,-\lambda)$, and may be written as
\begin{equation}\label{gcs}
    g(\tau,\lambda)=\ns\,2\tau\,\ns\,2\lambda\,\ns\, 2(\tau+\lambda)
    [1-\cn\,2\tau\,\cn\,2\lambda\,\cn\, 2(\tau+\lambda)].
\end{equation}
For a particular case $\lambda={\rm {\bf K}}/2$, to be
important for non-periodic limit,
\begin{eqnarray}\label{calC}
    &g\left(\tau,\frac{1}{2}{\rm {\bf K}}\right)=\mathcal{C}(\tau)\,,\qquad
    \mathcal{C}(\tau)=\ns\,2\tau\,\nc\, 2\tau\, \dn\, 2\tau
    \,.&
\end{eqnarray}
Notice that $g(\tau,\lambda)$ takes nonzero values for all
real values of its arguments~\footnote{It takes zero values
at some complex values of the arguments, for instance,
$\mathcal{C}(\frac{1}{2}{\rm {\bf K}}\pm\frac{i}{2}{\rm
{\bf K}}')=0$.}. Equation  (\ref{Deladd}) is a kind of
addition formula for elliptic function $\Delta(x;\tau)$.
Differentiating (\ref{Deladd}) in $x$ and using Ricatti
equations (\ref{Ricatti}), we obtain a relation
\begin{eqnarray}
    \Delta'(x+\tau+\lambda;\lambda)-
    \Delta(x+\lambda;\tau+\lambda)\Delta(x+\tau+\lambda;\lambda)=\nonumber\\
    -\frac{1}{2}\left(\Delta^2(x;\tau)+\Delta'(x;\tau)+
    \delta_2(\tau)\right)
    -g(\tau,\lambda)\Delta(x;\tau)+G(\tau,\lambda)\,,
    \label{Dcal}
\end{eqnarray}
where $G(\tau,\lambda)=\frac{1}{2}
\left[1+k^2+g^2(\tau,\lambda)-\ns^2\, 2\tau\, - \ns^2\,
2\lambda-\ns^2\, 2(\tau+\lambda)\right]\equiv 0$.

In conclusion of this section we note that functions
$\delta_a(\tau)$, $a=0,1,2$, can be given a physical sense
by expressing them in terms of the band edges energies and
of $\varepsilon(\tau)$\,: $
    \delta_2(\tau)=-(\tilde{E}_1^{2}+\tilde{E}_2^{2}+
    \tilde{E}_3^{2}),
$
$
    \delta_1(\tau)=-2\frac{d\tilde{E}_1}{d\tau},
$
$
    \delta_0(\tau)=-\delta_2(\tau)
    -2(\tilde{E}_1\tilde{E}_2+\tilde{E}_1\tilde{E}_3
    +\tilde{E}_2\tilde{E}_3),
$ where $\tilde{E}_i(\tau)=E_i+\varepsilon(\tau)$, $E_1=0$,
$E_2=k'{}^2$ and $E_3=1$. Particularly, $\delta_1$ measures
a velocity with which a scale of supersymmetry breaking
changes as a function of the shift parameter. Notice also
that the first equation in (\ref{DelEq23}) has a form of a
modified Ginzburg-Landau equation, see \cite{CofShLL},
which corresponds here to a gap equation for the real
condensate field in the kink-antikink crystalline phase in
the Gross-Neveu model with a bare mass term, see
\cite{Thies2,BasDun2}. At $\tau=(\frac{1}{2}+n){\rm {\bf
K}}$  we have $\delta_1=0$, and superpotential $\Delta(x)$
satisfies the nonlinear Schrodinger equation,  the lowest
nontrivial member of the modified Korteweg-de Vries
hierarchy \cite{GezHol}. This homogenisation of the second
order nonlinear differential equation can be associated
with restoration of the discrete chiral symmetry in
(\ref{LagGN}) at $m_0=0$.

\section{Higher order integrals and nonlinear superalgebra}\label{highinter}

Now we are in a position to identify higher order local
intertwining operators and integrals of motion for the
system $\mathcal{H}$. First, we find the second order
intertwining operators. Changing $\tau\rightarrow -\lambda$
and shifting the argument $x\rightarrow x+\tau+\lambda$ in
the first relation from (\ref{DHinter}), we obtain
\begin{equation}\label{Htlam}
    \mathcal{D}(x+\tau+\lambda;-\lambda)H(x+\tau)=
    H(x+\tau+2\lambda)\mathcal{D}(x+\tau+\lambda;-\lambda).
\end{equation}
Multiplying (\ref{Htlam}) by
$\mathcal{D}(x+\lambda;\tau+\lambda)$ from the left, and
using once again (\ref{DHinter}) on the right hand side, we
obtain an intertwining relation
\begin{equation}\label{AcalH}
    \mathcal{B}(x;\tau,\lambda)H(x_+)=
    H(x_-)\mathcal{B}(x;\tau,\lambda)\,.
\end{equation}
It is generated by the second order differential operator
\begin{equation}\label{Acal}
    \mathcal{B}(x;\tau,\lambda)=
    \mathcal{D}(x+\lambda;\tau+\lambda)
    \mathcal{D}^\dagger(x+\tau+\lambda;\lambda)\,,
\end{equation}
which is defined for $\lambda\,,\tau+\lambda\neq n{\rm {\bf
K}}$. For adjoint operator we have
$\mathcal{B}^\dagger(x;\tau,\lambda)H(x-\tau)=
H(x+\tau)\mathcal{B}^\dagger(x;\tau,\lambda)$. In
accordance with (\ref{Htlam}),  the second order
intertwining operator (\ref{Acal})  shifts the
Hamiltonian's argument first for $2\lambda$ and then for
$-2(\tau+\lambda)$. Equivalent representation of the
operator (\ref{Acal}) is
\begin{equation}\label{Aexpl}
    \mathcal{B}(x;\tau,\lambda)=
    -\mathcal{Y}(x;\tau)-g(\tau,\lambda)\mathcal{D}(x;\tau)\,,
\end{equation}
\begin{equation}\label{Ytau}
    \mathcal{Y}(x;\tau)=
    \frac{d^2}{dx^2}-\Delta(x;\tau)\frac{d}{dx}
    -\frac{1}{2}\left(
    \Delta^2(x;\tau)+\Delta'(x;\tau)+\delta_2(\tau)\right)\,,
    \qquad
    \mathcal{Y}^\dagger(x;\tau)=\mathcal{Y}(x;-\tau)\,.
\end{equation}
We have used here Eq. (\ref{Deladd}). So, the dependence of
$\mathcal{B}(x;\tau,\lambda)$ on $\lambda$ is localized
only in the $x$-independent multiplier $g(\tau,\lambda)$,
see Eq. (\ref{gcs}).

{}From Eqs. (\ref{Acal}) and  (\ref{DHinter}) it follows
that at $\tau=0$ the second order intertwining operators
$\mathcal{B}(x;\tau,\lambda)$ and
$\mathcal{B}^\dagger(x;\tau,\lambda)$ reduce, up to an
additive term $\varepsilon(\lambda)$, to the isospectral
superpartner Hamiltonians,
$
    \mathcal{B}(x;0,\lambda)=
    H(x)+\varepsilon(\lambda)
$,~\footnote{One could conclude that Eq. (\ref{Aexpl}) contradicts
to this relation since $g(\tau,\lambda)$ diverges at $\tau=0$, and
operators $\mathcal{D}(x;\tau)$ and $\mathcal{Y}(x;\tau)$ are not
defined for $\tau=0$. Eq. (\ref{Aexpl}) correctly  reproduces this
relation by treating $\tau=0$ as a limit $\tau\rightarrow 0$, and
employing addition formulae (\ref{addJs}) for Jacobi elliptic
functions.} $
    \mathcal{B}^\dagger(x;0,\lambda)=H(x+2\lambda)+\varepsilon(\lambda).
$

Forgetting for the moment on the $\tau=0$ case, from the
viewpoint of intertwining relation (\ref{AcalH}), one could
conclude that the parameter $\lambda$ has a ``gauge-like",
non-observable nature. Such a conclusion, however,  is not
correct. We will return to this point later.

Since $g(\tau,\lambda)$ is nonzero for real $\tau$ and
$\lambda$, operator $\mathcal{Y}(x;\tau)$, unlike
$\mathcal{B}(x;\tau,\lambda)$, is not factorizable in terms
of our first order intertwining operators (with real shift
parameters)~\footnote{It can be factorized in terms of our
first order Darboux operators $\mathcal{D}$ in special
cases of $\tau=(\frac{1}{2}+n){\rm {\bf K}}$. Such a
factorization corresponds  to complex values of the shift
parameters, see a discussion below in this section.}.
Nevertheless, it is the second order intertwining operator
as well as $\mathcal{B}(x;\tau,\lambda)$. It can be
presented as a linear combination of the second and first
order intertwining operators,
$\mathcal{Y}(x;\tau)=-\mathcal{B}(x;\tau,\lambda)-
g(\tau,\lambda)\mathcal{D}(x;\tau)$, and also may be used
together with the first order operator
$\mathcal{D}(x;\tau)$ to characterize the system. At the
end of this section we shall discuss the peculiarities
associated with such an alternative.

Having in mind a non-periodic limit we discuss later, it is
convenient to fix $\lambda={\rm {\bf K}}/2$, and introduce
a notation $\mathcal{A}(x;\tau)=
    \mathcal{B}\left(x;\tau,\frac{1}{2}{\rm {\bf
K}}\right)$,
 i.e.
\begin{eqnarray}\label{Axt}
    &\mathcal{A}(x;\tau)=\mathcal{D}
    \left(x+\frac{1}{2}{\rm {\bf K}};\tau+\frac{1}{2}{\rm {\bf K}}
    \right)\mathcal{D}^\dagger
    \left(x+\tau+\frac{1}{2}{\rm {\bf K}};\frac{1}{2}{\rm {\bf K}}
    \right)=-\mathcal{Y}(x;\tau)-
    \mathcal{C}(\tau)\mathcal{D}(x;\tau)\,,&
\end{eqnarray}
where $\mathcal{C}(\tau)$ is defined in Eq. (\ref{calC}).
Employing the properties of $\mathcal{Y}(x;\tau)$ and
$\mathcal{D}(x;\tau)$ under hermitian conjugation, from
(\ref{Axt}) one finds
$\mathcal{A}^\dagger(x;\tau)=\mathcal{A}(x;-\tau)$, and
then a representation alternative to (\ref{Axt}) is
obtained,
$
    \mathcal{A}(x;\tau)=\mathcal{D}(x-\tau+\frac{1}{2}{\rm {\bf K}};
    \frac{1}{2}{\rm {\bf K}})
    \mathcal{D}^\dagger(x+\frac{1}{2}{\rm {\bf K}};-\tau
    +\frac{1}{2}{\rm {\bf K}})$.
 Unlike the operators
$\mathcal{D}(x;\tau)$ and $\mathcal{Y}(x;\tau)$, the
$\mathcal{A}(x;\tau)$ is well defined at $\tau=0$ and
reduces just to a non-shifted Hamiltonian, $
\mathcal{A}(x;0)=\mathcal{A}^\dagger(x;0)=H(x)$. Notice,
however, that unlike $\mathcal{D}(x;\tau)$, it is not
defined for $\tau=(\frac{1}{2}+n){\rm {\bf K}}$.

Second order intertwining operator of the most general form
(\ref{Acal}) may be presented in terms of the intertwining
operators $\mathcal{A}(x;\tau)$ and $\mathcal{D}(x;\tau)$,
$
    \mathcal{B}(x;\tau,\lambda)=\mathcal{A}(x;\tau)+
    \big(C(\tau)-g(\tau,\lambda)\big)
    \mathcal{D}(x;\tau).
$

Because of Eq. (\ref{AcalH}), the self-isospectral system
possesses (for $\tau\neq (\frac{1}{2}+n){\rm {\bf K}}$) the
second order integrals
\begin{equation}\label{Q}
    Q_1=\left(%
    \begin{array}{cc}
  0 & \mathcal{A}^\dagger(x;\tau) \\
      \mathcal{A}(x;\tau) & 0 \\
    \end{array}%
    \right),\qquad
    Q_2=i\sigma_3Q_1\,
\end{equation}
to be nontrivial for $\tau\neq n{\rm {\bf K}}$ and
independent from the first order integrals (\ref{Sdef}).

With some algebraic manipulations, we find
\begin{equation}\label{AA+}
    \mathcal{A}^\dagger(x;\tau)\mathcal{A}(x;\tau)=
    H(x_+)\left[H(x_+)+\varrho(\tau)\right]\,,\quad
    {\rm where}\quad
    \varrho(\tau)=k'{}^2\sn^2\,2\tau\,\nc^2\,2\tau\,.
\end{equation}
 A similar relation is obtained from (\ref{AA+}) by a simple change
$\tau\rightarrow-\tau$, $
    \mathcal{A}(x;\tau)\mathcal{A}^\dagger(x;\tau)=
    H(x_-)\left[H(x_-)+\varrho(\tau)\right].
$
cf.  relations in (\ref{Hx+t}) for the first order
intertwining operators.

The intertwining second order operator
$\mathcal{A}(x;\tau)$ annihilates the lower energy state
$\dn\,(x+\tau)$ of the system $H(x+\tau)$. Another state
annihilated by it is
\begin{equation}\label{fxtnon}
    f(x,\tau)=\dn\,(x+\tau)\int^x
    \frac{F(u+\frac{1}{2}{\rm {\bf K}};\tau+\frac{1}{2}{\rm {\bf K}})}
    {\dn\,(u+\tau)}du\,,
\end{equation}
and we have $f(x+2K,\tau)=\exp\left[2{\rm {\bf K}}{\rm
z}(\tau+\frac{1}{2}{\rm {\bf K}})\right]\,f(x,\tau)$.
Function (\ref{fxtnon}) for $\tau\neq 0$ is unbounded and
describes therefore  a non-physical eigenstate of
$H(x+\tau)$ from the lower forbidden band with energy
$E=-\varrho(\tau)<0$, see Eq. (\ref{AA+}). At $\tau=0$,
function (\ref{fxtnon}) reduces to ${\rm E}(x+{\rm {\bf
K}})\,\dn\,x$ that corresponds to a nonphysical state of
$H(x)$ of zero eigenvalue.

Like the first order operator $\mathcal{D}(x;\tau)$,
$\mathcal{A}(x;\tau)$ transforms the eigenstates of
$H(x+\tau)$ into those of $H(x-\tau)$,
\begin{equation}\label{Aintersusy}
    \mathcal{A}(x;\tau)\Psi^\alpha_\pm(x_+)
    =\mathcal{F}^{\mathcal{A}}_\pm(\alpha,\tau)\,
    \Psi^\alpha_\pm(x_-)\,,
\end{equation}
where
\begin{equation}\label{FA}
    \mathcal{F}^{\mathcal{A}}_\pm(\alpha,\tau)=e^{\pm i\varphi^{\mathcal{A}}(\alpha,\tau)}
    \mathcal{M}^{\mathcal{A}}(\alpha,\tau)\,,\quad
    \mathcal{M}^{\mathcal{A}}(\alpha,\tau)=\sqrt{E(\alpha)(E(\alpha)+
    \varrho(\tau))}\,.
\end{equation}
The modulus and the phase of the complex amplitude
$\mathcal{F}^{\mathcal{A}}_\pm(\alpha,\tau)$ are expressed
in terms of those for the first order intertwining operator
by employing Eqs. (\ref{Htlam}), (\ref{Axt}) and
(\ref{Dintersusy}),
\begin{eqnarray}\label{ADamphase}
    &\mathcal{M}^{\mathcal{A}}(\alpha,\tau)=
    \mathcal{M}^{\mathcal{D}}\left(\alpha,\tau+\frac{1}{2}{\rm {\bf K}}\right)
    \mathcal{M}^{\mathcal{D}}\left(\alpha,\frac{1}{2}{\rm {\bf K}}\right)\,,\quad
    \varphi^{\mathcal{A}}(\alpha,\tau)=\varphi^{\mathcal{D}}
    (\alpha,\tau+\frac{1}{2}{\rm {\bf K}})-\varphi^{\mathcal{D}}(\alpha,
    \frac{1}{2}{\rm {\bf K}})\,.\qquad
    &
\end{eqnarray}
 A
phase $\varphi^{\mathcal{A}}(\alpha,\tau)\in\R$ has, unlike
(\ref{varphiD}), a property $
e^{i\varphi^{\mathcal{A}}(\alpha,-\tau)}=e^{-i\varphi^{\mathcal{A}}
    (\alpha,\tau)}$
due to a relation
$\mathcal{A}^\dagger(x;\tau)=\mathcal{A}(x;-\tau)$ to be
different in sign from that for the first order
intertwining operator,
$\mathcal{D}^\dagger(x;\tau)=-\mathcal{D}(x;-\tau)$. For
the edge band states, particularly, we have
$
    \mathcal{A}(x;\tau)\psi_i(x_+)=
    \mathcal{F}^{\mathcal{A}}_i(\tau)\psi_i(x_-)$,
$    \mathcal{A}^\dagger(x;\tau)\psi_i(x_-)=
    \mathcal{F}^{\mathcal{A}}_i(\tau)\psi_i(x_+),
$
where
$
    \mathcal{F}^{\mathcal{A}}_i(\tau)=
    0\,,\,\,k'{}^2\nc\,2\tau\,,\,\,
    \dn\,2\tau\,\nc\,2\tau$, $ i=1,2,3$, cf.
    (\ref{psiiCi}).
The eigenstates of the integral $Q_1$, see  (\ref{Q}), have
a form similar to that for $S_1$,
\begin{equation}\label{Q1eigen}
    Q_1\Psi^\alpha_{\pm,Q_1,\epsilon}=\epsilon
    \mathcal{M}^{\mathcal{A}}(\alpha,\tau)
    \Psi^\alpha_{\pm,Q_1,\epsilon}\,,\quad
    \Psi^\alpha_{\pm,Q_1,\epsilon}=
    \left(%
\begin{array}{c}
  \Psi^\alpha_\pm(x_+) \\
  \epsilon e^{\pm i\varphi^\mathcal{A}(\alpha,\tau)}
    \Psi^\alpha_\pm(x_-) \\
\end{array}%
\right),\quad \epsilon=\pm1\,.
\end{equation}

\vskip0.2cm

Two relations are valid for the first and second order intertwining
operators,
\begin{equation}\label{DAAD}
    \mathcal{D}^\dagger(x;\tau)\mathcal{A}(x;\tau)=
    \mathcal{P}(x_+)-\mathcal{C}(\tau)H(x_+)\,,\quad
    \mathcal{D}(x;\tau)\mathcal{A}^\dagger(x;\tau)=
    -\mathcal{P}(x_-)-\mathcal{C}(\tau)H(x_-)\,.
\end{equation}
 Here $\mathcal{P}(x_+)=\mathcal{P}(x+\tau)$ is an anti-hermitian
third order differential operator
\begin{eqnarray}
    \mathcal{P}(x_+)&=&\frac{d^3}{dx^3}-
    \frac{3}{2}\left(\Delta^2+\Delta'+
\frac{1}{3}\delta_2(\tau)\right)
    \frac{d}{dx}
    -\frac{3}{4}\left(\Delta^2+\Delta'\right)'\quad\nonumber\\
    &=&\frac{d^3}{dx^3}+
    \left(1+k^2-3k^2\sn^2x_+\right)\frac{d}{dx}
    -3k^2\sn\,x_+\,\cn\,x_+\,\dn\,x_+\,.\quad\label{Zx+t}
\end{eqnarray}
 Notice that like the Lam\'e Hamiltonian,
the operator (\ref{Zx+t}) is well defined for any value of
the shift parameter $\tau$. Two related equalities may be
obtained from (\ref{DAAD}) by hermitian conjugation.

 Making use of intertwining relations
(\ref{DHinter}), (\ref{AcalH}), we find that $H(x+\tau)$
commutes with
$\mathcal{D}^\dagger(x;\tau)\mathcal{A}(x;\tau)$, and,
therefore, $\mathcal{P}(x+\tau)$ is an integral for the
subsystem $H(x+\tau)$. For self-isospectral supersymmetric
system $\mathcal{H}$  we have then two further, third order
hermitian integrals
\begin{equation}\label{Zint}
    L_1=-i\,{\rm diag}\,\left(
  \mathcal{P}(x_+), \mathcal{P}(x_-)
    \right),\qquad
    L_2=\sigma_3 L_1\,.
\end{equation}
Operator $\mathcal{P}(x)$ is a Lax operator for the
periodic one-gap Lam\'e system $H(x)$, see \cite{CP2,Tri}.

The following relations that involve the operator
$\mathcal{P}(x_+)$ are valid,
\begin{equation}\label{DZ}
    \mathcal{D}(x;\tau)\mathcal{P}(x+\tau)=
    \mathcal{A}(x;\tau)\left[H(x_+)+\varepsilon(\tau)\right]
    +\mathcal{C}(\tau)\mathcal{D}(x;\tau)H(x_+)\,,
\end{equation}
\begin{equation}\label{AZ}
    \mathcal{A}(x;\tau)\mathcal{P}(x_+)=
    -\mathcal{D}(x;\tau)H(x_+)\left[H(x_+)+\varrho(\tau)\right]-
    \mathcal{C}(\tau)\mathcal{A}(x;\tau)
     H(x_+)\,,
\end{equation}
\begin{equation}\label{PP}
    -\mathcal{P}^2(x_+)=P(H(x_+))\,,\qquad
    P(H)=H(H-k'{}^2)(H-1)\,.
\end{equation}
The third order polynomial $P(H)$ is the same spectral
polynomial of the Lam\'e system that arose before in
(\ref{dEdb}) and in differential dispersion relation
(\ref{dkdE})\,: it turns into zero when acts on the edge
states with energies $E_i=0,k'{}^2,1$. Since the third
order differential operator $\mathcal{P}(x_+)$ is an
integral of motion for $H(x_+)$, relation (\ref{PP}) means
that the edge states $\dn\,x_+$, $\cn\,x_+$ and $\sn\,x_+$
form its kernel \cite{Tri}. The spectral polynomial is a
semi-positive definite operator, while $\mathcal{P}(x)$ is
an anti-hermitian operator. Its action on physical Bloch
states (\ref{Psipm}) should reduce therefore to $\pm
i\sqrt{P(E(\alpha))}$. The phase cannot change abruptly
within the allowed bands. To fix correctly the sign, one
can consider a limit $k\rightarrow 0$, in which Lam\'e
system (\ref{HLame}) reduces to a free particle, an
integral $\mathcal{P}(x)$ reduces to a third order operator
$d^3/dx^3 +d/dx$, forbidden zone $k'{}^2<E<1$ disappears,
Bloch states transform into the plane wave states, whereas
the edge states $\dn\,x$, $\cn\,x$ and $\sn\,x$ reduce,
respectively, to $1$, $\cos x$ and $\sin x$ with energies
$E=0,\,1$ and $1$. Summarizing all this, one finds that the
operator (\ref{Zx+t}) acts on the physical Bloch states
(\ref{Psipm}) as follows,
\begin{equation}\label{LaxBloch}
    \mathcal{P}(x)\Psi^\alpha_\pm(x)=\mp i
    \eta(E)\sqrt{P(E(\alpha))}\,
    \Psi^\alpha_\pm(x)\,,
\end{equation}
where, as in (\ref{dEdb}) and (\ref{dkdE}),  $\eta(E)=-1$ for
valence and $+1$ for conduction bands~\footnote{ Applying the first
relation from (\ref{DAAD}) to a physical Bloch state
$\Psi^\alpha_+(x_+)$ and using  an equality
$E(E+\varrho(\tau))(E+\varepsilon(\tau))=P(E)+\mathcal{C}^2(\tau)E^2$,
we obtain the Pythagorean relation for a rectangular triangle  with
legs $\mathcal{C}(\tau)E(\alpha)$ and $\sqrt{P(E(\alpha))}$, $
    \sqrt{P(E(\alpha))+\mathcal{C}^2(\tau)E^2(\alpha)}\,e^{i
    (\varphi^{\mathcal{D}}(\alpha,\tau+\frac{K}{2})-
    \varphi^{\mathcal{D}}(\alpha,\tau)-
    \varphi^{\mathcal{D}}(\alpha,\frac{K}{2}))}=i\eta
    \sqrt{P(E(\alpha))}+\mathcal{C}(\tau)E(\alpha).
$}. Relation (\ref{LaxBloch}) means, particularly, that the
Lax operator is not reduced just to a square root from the
spectral polynomial since Hamiltonian does not distinguish
index $\pm$. This is a true, nontrivial integral of motion
that is related with the Hamiltonian $H$ by polynomial
equation (\ref{PP})~\footnote{This corresponds to
Burchnall-Chaundy theorem \cite{BCI} that underlies the
theory of nonlinear integrable systems \cite{finegap}. It
asserts that if two ordinary differential in $x$ operators
$A$ and $B$ of mutually prime orders $l$ and $m$ do
commute, they obey a relation $P(A,B) = 0$, where $P$ is a
polynomial of order $m$ in $A$, and of order $l$ in $B$.}.
Eq. (\ref{PP}) corresponds to a non-degenerate spectral
elliptic curve of genus one associated with a one-gap
periodic Lam\'e system \cite{finegap}.

Let us discuss now the superalgebra generated by the zero,
$\sigma_3$, first, $S_a$, second, $Q_a$, and third, $L_a$,
order integrals of motion of the self-isospectral system
$\mathcal{H}$. The operator $\Gamma=\sigma_3$ commutes with
$L_a$ and anti-commutes with $Q_a$, and so, classifies
them, respectively,  as bosonic and fermionic operators.
Using the displayed relations for the operators
$\mathcal{D}$, $\mathcal{A}$ and $\mathcal{P}$ as well as
those obtained from them by a hermitian conjugation and by
a change $\tau\rightarrow -\tau$, Eq. (\ref{SSH}) is
extended by the anti-commutation relations of the integrals
$S_a$ with $Q_a$, and the commutation relations of $S_a$
and $Q_a$ with $L_a$. We arrive as a result at the
following superalgebra for the self-isospectral system
(\ref{Hcal}) with $\Z_2$ grading  operator
$\Gamma=\sigma_3$,
\begin{equation}\label{SSQQ}
    \{S_a,S_a\}=2\delta_{ab}\big(\mathcal{H}+\varepsilon(\tau)\big),\qquad
    \{Q_a,Q_b\}=2\delta_{ab}\mathcal{H}\big(
    \mathcal{H}+\varrho(\tau)\big),
\end{equation}
\begin{equation}\label{QS}
    \{S_a,Q_b\}=2\left(-\delta_{ab}\,\mathcal{C}(\tau)\mathcal{H}
    +\epsilon_{ab}\,L_1\right),
\end{equation}
\begin{equation}\label{SL2}
     [L_1,S_a]=[L_1,Q_a]=[L_1,L_2]=0\,,\qquad
    [L_2,S_a]=2i\big(S_a \mathcal{C}(\tau)\mathcal{H}
    +Q_a(\mathcal{H}+\varepsilon(\tau))
    \big)\,,
\end{equation}
\begin{equation}\label{QL2}
    [L_2,Q_a]=-2i\big(
    S_a\mathcal{H}(\mathcal{H}+\varrho(\tau))+Q_a \mathcal{C}(\tau)\mathcal{H}
    \big)\,,
\end{equation}
\begin{equation}\label{sigInt}
    [\sigma_3,S_a]=-2i\epsilon_{ab}S_b,\quad
    [\sigma_3,Q_a]=-2i\epsilon_{ab}Q_b,\quad
    [\sigma_3,L_a]=0,
\end{equation}
\begin{equation}\label{HcomInt}
    [\mathcal{H},\sigma_3]=[\mathcal{H},S_a]=
    [\mathcal{H},Q_a]=[\mathcal{H},L_a]=0\,.
\end{equation}
We have here a nonlinear superalgebra, in which $L_1$ (that
is a Lax operator for $\mathcal{H}$) plays a role of the
bosonic central charge, and $\sigma_3$ is treated as one of
its even generators in correspondence with $\Z_2$ grading
relations
$[\sigma_3,\sigma_3]=[\sigma_3,\mathcal{H}]=[\sigma_3,L_a]=0$
and $\{\sigma_3,S_a\}=\{\sigma_3,Q_a\}=0$.

Since $L_1$ commutes with $S_a$ and $Q_a$, the eigenstates
(\ref{S1eigen}) and (\ref{Q1eigen})  of $S_1$ and $Q_1$ are
simultaneously the eigenstates of $L_1$,
\begin{equation}\label{L1eigen}
    L_1\Psi^\alpha_{\pm,\Lambda,\epsilon}=
    \mp\eta\sqrt{P(\alpha)}\,
    \Psi^\alpha_{\pm,\Lambda,\epsilon}\,,
\end{equation}
where $\Lambda=S_1$ or $Q_1$, $\eta$ is the same as in
(\ref{dkdE}) and (\ref{LaxBloch}), and
$P(\alpha)=P(E(\alpha))$. Note that unlike $S_1$ and $Q_1$,
$L_1$ distinguishes the index $\pm$.

In correspondence with the discussion related to
(\ref{fxtnon}), the $Q_a$, $a=1,2$, annihilate the two
ground states of zero energy, $\dn\,(x+\tau)$ and
$\dn\,(x-\tau)$, while other two states from their kernel
are non-physical. These supercharges are not defined,
however, for $\tau=(\frac{1}{2}+n){\rm {\bf K}}$, which are
the only values of the shift parameter when the $N=2$
supersymmetry associated with the first order supercharges
$S_a$ is not broken. Therefore, when the first and the
second order supercharges are simultaneously defined [for
$\tau\neq (\frac{1}{2}+n){\rm {\bf K}},\, n{\rm {\bf K}}$],
the supersymmetry generated together by $S_a$ and $Q_a$ is
partially broken.

One could construct, instead, the second order
supercharges, $Q^{\mathcal{Y}}_a$,  on the basis of the
intertwining operators $\mathcal{Y}(x;\tau)$ and
$\mathcal{Y}^\dagger(x;\tau)$. According to (\ref{Axt}),
they are related to $Q_a$ as
\begin{equation}\label{QY}
    Q^{\mathcal{Y}}_a=-Q_a-\mathcal{C}(\tau)S_a\,.
\end{equation}
The corresponding super-algebra with $Q_a$ substituted for
$Q^{\mathcal{Y}}_a$ will have then a form similar to that
we have discussed, with the change of some of the
corresponding  (anti)-commutators for
\begin{equation}\label{QYsusy1}
    \{Q^{\mathcal{Y}}_a,Q^{\mathcal{Y}}_b\}=2\delta_{ab}\big(\mathcal{H}(
    \mathcal{H}+\varrho(\tau)-\mathcal{C}^2(\tau))+
    \varepsilon(\tau)\mathcal{C}^2(\tau)\big),
\end{equation}
\begin{equation}\label{QYsusy2}
     \{S_a,Q^{\mathcal{Y}}_b\}=-2\big(\delta_{ab}\,\sigma_3\,\mathcal{C}(\tau)
    \varepsilon(\tau)+
    \epsilon_{ab}\,L_1\big)\,,
\end{equation}
\begin{equation}\label{QYsusy3}
    [L_2,S_a]=-2i\big(
    S_a\mathcal{C}(\tau)\varepsilon(\tau)
    +Q^{\mathcal{Y}}_a(\mathcal{H}+\varepsilon(\tau))
    \big)\,,
\end{equation}
\begin{equation}\label{QYsusy4}
    [L_2,Q^{\mathcal{Y}}_a]=2i\big(
    S_a\mathcal{H}(\mathcal{H}+\varrho(\tau)+\varepsilon(\tau)
    \mathcal{C}(\tau)-\mathcal{C}^2(\tau))+Q^{\mathcal{Y}}_a\varepsilon(\tau)
    \mathcal{C}(\tau)
    \big)\,.
\end{equation}
The second order supercharges $Q^{\mathcal{Y}}_a$, like
$S_a$, are well defined at $\tau=(\frac{1}{2}+n){\rm {\bf
K}}$ but not defined for $\tau=n{\rm {\bf K}}$. Analyzing
the roots of the polynomial in the right hand side of
(\ref{QYsusy1}), one finds that the kernels of
$Q^{\mathcal{Y}}_a$, $a=1,2$, for
$\tau\neq(\frac{1}{2}+n){\rm {\bf K}}$ are formed by
non-physical states. In the exceptional  case
$\tau=(\frac{1}{2}+n){\rm {\bf K}}$, for which the
supercharges $Q_a$ are not defined, the polynomial in
(\ref{QYsusy1}) reduces  to the second order polynomial
\begin{equation}\label{PQYH}
    P_{Q^{\mathcal{Y}}}(\mathcal{H})=
    (\mathcal{H}-k'{}^2)(\mathcal{H}-1)\,.
\end{equation}
In correspondence with this, the zero modes of the
operators $\mathcal{Y}(x;\frac{1}{2}{\rm {\bf K}})$ and
$\mathcal{Y}^\dagger(x;\frac{1}{2}{\rm {\bf
K}})=\mathcal{Y}(x;-\frac{1}{2}{\rm {\bf K}})$ are,
respectively, the physical edge states
$\cn\,(x+\frac{1}{2}{\rm {\bf K}})$,
$\sn\,(x+\frac{1}{2}{\rm {\bf K}})$ and
$\cn\,(x-\frac{1}{2}{\rm {\bf K}})$,
$\sn\,(x-\frac{1}{2}{\rm {\bf K}})$. This property reflects
a peculiarity of the case $\tau=(\frac{1}{2}+n){\rm {\bf
K}}$ in another aspect. In accordance with footnote 5,
function $g(\tau,\lambda)$ in (\ref{Aexpl}) turns into zero
at $\lambda=\frac{1}{2}({\rm {\bf K}}+i{\rm {\bf K}}')$.
The second order operator $\mathcal{Y}(x;\frac{1}{2}{\rm
{\bf K}})$ factorizes then either as
$\mathcal{Y}(x;\frac{1}{2}{\rm {\bf
K}})=-\mathcal{D}(x+\frac{1}{2}({\rm {\bf K}}+i{\rm {\bf
K}}');{\rm {\bf K}}+\frac{1}{2}i{\rm {\bf K}}')
\mathcal{D}^\dagger(x+{\rm {\bf K}}+\frac{1}{2}i{\rm {\bf
K}}';\frac{1}{2}({\rm {\bf K}}+i{\rm {\bf K}}'))$, or in
alternative  form obtained by the change of $i$ for $-i$.
These two factorizations can be presented equivalently as
\begin{eqnarray}\label{Ysncn1}
    &\mathcal{Y}(x;\frac{1}{2}{\rm {\bf
    K}})=\left(\ns\,(x-\frac{1}{2}{\rm {\bf
    K}})\frac{d}{dx}\sn\,(x-\frac{1}{2}{\rm {\bf
    K}})\right)\left(\cn\,(x+\frac{1}{2}{\rm {\bf
    K}})\frac{d}{dx}\nc\,(x+\frac{1}{2}{\rm {\bf
    K}})\right)\,,&\\
    &\mathcal{Y}(x;\frac{1}{2}{\rm {\bf
    K}})=\left(\nc\,(x-\frac{1}{2}{\rm {\bf
    K}})\frac{d}{dx}\cn\,(x-\frac{1}{2}{\rm {\bf
    K}})\right)\left(\sn\,(x+\frac{1}{2}{\rm {\bf
    K}})\frac{d}{dx}\ns\,(x+\frac{1}{2}{\rm {\bf
    K}})\right)\,.&\label{Ysncn2}
\end{eqnarray}

{}From here we see
that the particular case of the half period shift of the
super-partner systems is indeed exceptional. In this case
not only the $N=2$ supersymmetry associated with the first
order supercharges $S_a$ is unbroken (when zero modes of
$S_a$ are the ground states that form a zero energy
doublet), but all the other edge states of the energy
doublets with $E=k'{}^2$ and $E=1$ correspond to zero modes
of the second order supercharges $Q^{\mathcal{Y}}_a$. Then
the third order spectral polynomial
$P(\mathcal{H})=\mathcal{H}(\mathcal{H}-k'{}^2)
(\mathcal{H}-1)$ is just a product of the first and the
second order polynomials which correspond to the squares of
the first, $S_a$, and the second, $Q^{\mathcal{Y}}_a$,
order supercharges. In this special case the
(anti-)commutation relations (\ref{QYsusy2}),
(\ref{QYsusy3}), (\ref{QYsusy4}) also simplify their form,
$
     \{S_a,Q^{\mathcal{Y}}_b\}=-2
    \epsilon_{ab}\,L_1, $
$
    [L_2,S_a]=-2iQ^{\mathcal{Y}}_a\mathcal{H},
$
$
    [L_2,Q^{\mathcal{Y}}_a]=2i
    S_aP_{Q^{\mathcal{Y}}}(\mathcal{H}).
$ We also have
\begin{equation}\label{S*Q=L}
    S_aQ^{\mathcal{Y}}_a=-Q^{\mathcal{Y}}_aS_a=
    -iL_2\,,\qquad
    S_aQ^{\mathcal{Y}}_b=Q^{\mathcal{Y}}_bS_a=-L_1\,,
\end{equation}
where there is no summation in index $a$, and $b\neq a$. This is in
conformity with the above mentioned factorization of the spectral
polynomial. However, since $Q^{\mathcal{Y}}_a$ does not annihilate
the ground states $\dn\,(x+\frac{1}{2}{\rm {\bf K}})$ and
$\dn\,(x-\frac{1}{2}{\rm {\bf K}})$ (which are transformed mutually
by the intertwining operators $\mathcal{Y}(x;\frac{1}{2}{\rm {\bf
K}})$ and $\mathcal{Y}^\dagger(x;\frac{1}{2}{\rm {\bf K}})$),  we
conclude that nonlinear supersymmetry of the self-isospectral system
also is partially broken at $\tau=(\frac{1}{2}+n){\rm {\bf
K}}$~\footnote{ Cf. this picture as well as that for $\tau\neq
(\frac{1}{2}+n){\rm {\bf K}}$, which  we discussed above, with the
picture of supersymmetry breaking in the systems with topologically
nontrivial Bogomolny-Prasad-Sommerfield states \cite{BPS}.}.
\vskip0.05cm

 In the next section we will see that
another peculiarity of our self-isospectral system is that
the choice $\Gamma=\sigma_3$ is not unique for
identification of the $\Z_2$ grading operator\,: it also
admits other choices for $\Gamma$, which lead to different
identifications of the integrals $\sigma_3$, $S_a$, $Q_a$
and $L_a$ as bosonic and fermionic operators. This results
in the alternative forms for the superalgebra. Each of such
alternative forms of the superalgebra makes, particularly,
a nontrivial relation (\ref{PP}) to be `visible' explicitly
just in its structure, unlike the case with
$\Gamma=\sigma_3$ that we have discussed. We also will
identify the integrals of motion which  detect the phases
in the structure of the  eigenstates of the operators $S_a$
and $Q_a$.

\section{Nonlocal $\Z_2$ grading operators}\label{NonlocalZ2}

 Let us introduce the operators of
reflection in $x$ and $\tau$,
$
    \mathcal{R}x\mathcal{R}=-x$,
$
    \mathcal{R}\tau\mathcal{R}=\tau$,
$
    \mathcal{R}^2=1$,
$
     \mathcal{T}\tau\mathcal{T}=-\tau$,
$
    \mathcal{T}x\mathcal{T}=x$,
$
    \mathcal{T}^2=1$.
They intertwine the superpartner Hamiltonians,
$\mathcal{R}H(x_+)=H(x_-)\mathcal{R}$,
$\mathcal{T}H(x_+)=H(x_-)\mathcal{T}$, and we find that the
self-isospectral supersymmetric system (\ref{Hcal})
possesses the hermitian integrals of motion
\begin{equation}\label{RTint}
    \mathcal{R}\sigma_1,\quad
    \mathcal{T}\sigma_1,\quad
    \mathcal{R}\sigma_2,\quad
    \mathcal{T}\sigma_2,\quad
    \mathcal{RT}\sigma_3,\quad
    \mathcal{RT}\,.
\end{equation}
Like for $\sigma_3$, a square of each of them equals 1.
{}From relations
\begin{eqnarray}\label{RDAP1}
    &\mathcal{R}\mathcal{D}(x;\tau)=
    \mathcal{D}^\dagger(x;\tau)\mathcal{R}\,,\quad
    \mathcal{R}\mathcal{A}(x;\tau)=
    \mathcal{A}^\dagger(x;\tau)\mathcal{R}\,,\quad
    \mathcal{R}\mathcal{P}(x_+)=-\mathcal{P}(x_-)\mathcal{R}\,,&\\
    \label{RDAP2}
    &\mathcal{T}\mathcal{D}(x;\tau)=
    -\mathcal{D}^\dagger(x;\tau)\mathcal{T}\,,\quad
    \mathcal{T}\mathcal{A}(x;\tau)=
    \mathcal{A}^\dagger(x;\tau)\mathcal{T}\,,\quad
    \mathcal{T}\mathcal{P}(x_+)=\mathcal{P}(x_-)\mathcal{T}\,,&
\end{eqnarray}
it follows that  $\mathcal{R}$ and $\mathcal{T}$ intertwine
also  the operators of the same order within the pairs
($\mathcal{D}(x;\tau)$, $\mathcal{D}^\dagger(x;\tau)$),
($\mathcal{A}(x;\tau)$, $\mathcal{A}^\dagger(x;\tau)$), and
($\mathcal{P}(x_+)$, $\mathcal{P}(x_-)$). As a result, each
of the nonlocal in $x$ or $\tau$, or in both of them,
integrals of motion (\ref{RTint}) either commutes or
anti-commutes with each of the nontrivial local integrals
$S_a$, $Q_a$ and $L_a$. Then each integral from
(\ref{RTint}) also may be chosen as the $\Z_2$ grading
operator for the self-isospectral system (\ref{Hcal}).
Corresponding $\Z_2$ parities together with those
prescribed by a local integral $\sigma_3$ are shown in
Table \ref{T1}. $\Z_2$ parities of the second order
integrals $Q^{\mathcal{Y}}_a$, defined in (\ref{QY}), are
also displayed; the equality
$\mathcal{C}(-\tau)=-\mathcal{C}(\tau)$ has to be employed
in their computation. Notice that $Q^{\mathcal{Y}}_a$,
$a=1,2$, always has the same $\Z_2$ parity as the $Q_a$
with the same value of the index $a$.

\begin{table}[ht]
\caption{$\Z_2$ parities of the local integrals.
}\label{T1}
\begin{center}
\begin{tabular}{|c||c|c|c|c|c|c|c|}\hline
$\Gamma$ & $\sigma_3$ & $S_1$ & $S_2$ & $Q_1$,
$Q^{\mathcal{Y}}_1$ & $Q_2$, $Q^{\mathcal{Y}}_2$ & $L_1$ &
$L_2$
\\\hline\hline
$\sigma_3$ & $+$ & $-$ & $-$ & $-$ & $-$ & $+$ & $+$
\\[1pt]\hline
$\mathcal{R}\sigma_1$ & $-$ & $+$ & $-$ & $+$ & $-$ & $-$ &
$+$
\\[1pt]\hline
$\mathcal{T}\sigma_1 $ & $-$ & $-$ & $+$ & $+$ & $-$ & $+$
& $-$
\\[1pt]\hline
$\mathcal{R}\sigma_2$ & $-$ & $-$ & $+$ & $-$ & $+$ & $-$ &
$+$
\\[1pt]\hline
$\mathcal{T}\sigma_2$ & $-$ & $+$ & $-$ & $-$ & $+$ & $+$ &
$-$
\\[1pt]\hline
$\mathcal{R}\mathcal{T}\sigma_3$ & $+$ & $+$ & $+$ & $-$ &
$-$ & $-$ & $-$
\\[1pt]\hline
$\mathcal{R}\mathcal{T}$ & $+$ & $-$ & $-$ & $+$ & $+$ &
$-$ & $-$
\\[1pt]\hline
\end{tabular}
\end{center}
\end{table}
A positive $\Z_2$ parity is assigned for the Hamiltonian
$\mathcal{H}$ by any of the integrals (\ref{RTint}). Then
for any choice of the grading operator presented in  Table
\ref{T1}, four of the eight local integrals $\sigma_3$,
$\mathcal{H}$, $S_a$, $L_a$ and $Q_a$ or
$Q^{\mathcal{Y}}_a$ are identified as bosonic
 generators, and four are identified as fermionic
generators of the corresponding nonlinear superalgebra. The
superalgebra may be found for each choice of $\Gamma$ from
the set of integrals (\ref{RTint}) by employing  the
quadratic products of the operators $\mathcal{D}$,
$\mathcal{A}$ and $\mathcal{P}$ that have been discussed in
the previous section. Alternatively, some of the
(anti)-commutators may be obtained with the help of the
already known (anti)-commutation relations and relations
between the generators that involve $\sigma_3$. For
instance, $[S_1,Q_1]=i\sigma_3\{S_1,Q_2\}$. As an example,
we display the explicit form of the superalgebraic
relations for the choice $\Gamma=\mathcal{RT}$,
\begin{equation}\label{RTs1}
    \{S_a,S_b\}=2\delta_{ab}\big(
    \mathcal{H}+\varepsilon(\tau)\big)\,,\quad
    \{S_a,L_1\}=2\epsilon_{ab}\big(
    Q_b(\mathcal{H}+\varepsilon(\tau))+\mathcal{C}(\tau)
    S_b\mathcal{H}\big)\,,\quad
    \{S_a,L_2\}=0\,,
\end{equation}
\begin{equation}\label{RTs2}
   \{L_1,L_1\}=\{L_2,L_2\}=2P(\mathcal{H}),\qquad
    \{L_1,L_2\}=2\sigma_3P(\mathcal{H}),
\end{equation}
\begin{equation}\label{RTs3}
    [Q_a,S_b]=2i\big(-\delta_{ab}L_2+\epsilon_{ab}\mathcal{C}(\tau)
    \sigma_3\mathcal{H}\big),\qquad
    [Q_1,Q_2]=-2i\sigma_3\mathcal{H}\big(
    \mathcal{H}+\varrho(\tau)\big),
\end{equation}
\begin{equation}\label{RTs4}
    [Q_a,L_1]=0,\qquad [Q_a,L_2]=2i\big(
    \mathcal{C}(\tau)Q_a\mathcal{H}+S_a\mathcal{H}
    (\mathcal{H}+\varrho(\tau))\big),
\end{equation}
which should be supplied by the commutation relations
(\ref{sigInt}) and (\ref{HcomInt}). $P(\mathcal{H})$ in
(\ref{RTs2}) is the spectral polynomial, see (\ref{PP}).

A fundamental polynomial relation (\ref{PP}) between the
Lax operator and the Hamiltonian, that underlies a very
special, finite-gap nature of Lam\'e system~\footnote{In a
generic situation the spectrum of a one-dimensional
periodic system has infinitely many gaps \cite{finegap}.},
does not show up in the superalgebraic structure for a
usual choice of the diagonal matrix $\sigma_3$ as the
grading operator $\Gamma$, but is involved explicitly in
the superalgebra in the form of the anticommutator of one
or both generators $L_a$, $a=1,2$, when any of six
non-local integrals (\ref{RTint}) is identified as
$\Gamma$.

Note that for $\Gamma=\mathcal{RT}$ as well as for any
other choice of the grading operator that involves the
operator $\mathcal{T}$, the constant $\mathcal{C}(\tau)$
anticommutes with the grading operator and should be
treated as an odd generator of the superalgebra. As a
result, the right hand side in the second anticommutator in
(\ref{RTs1}) is an even operator, while the right hand side
in the first (second) commutator in (\ref{RTs3}) (in
(\ref{RTs4})) is an odd operator as it should be.

By employing Eq. (\ref{QY}), one can rewrite superalgebraic
relations (\ref{RTs1}), (\ref{RTs3}) and (\ref{RTs4}) in
terms of the integrals $Q^{\mathcal{Y}}_a$, which, unlike
$Q_a$, are defined for $\tau=(\frac{1}{2}+n){\rm {\bf K}}$.
We do not display them here, but write down only a
commutation relation
\begin{equation}\label{AQYeps}
    [S_a,Q^{\mathcal{Y}}_b]=2i\big(\delta_{ab}L_2+
    \sigma_3\epsilon_{ab}\mathcal{C}(\tau)
    \varepsilon(\tau)\big)\,,
\end{equation}
which we will need below. The form of such a superalgebra
simplifies significantly at $\tau=(\frac{1}{2}+n){\rm {\bf
K}}$ in correspondence with a special nature that the
integrals $S_a$ and $Q_a^{\mathcal{Y}}$ acquire in the
case. Particularly,  one finds
\begin{equation}\label{tK2QY1}
    \{S_a,S_b\}=2\delta_{ab}\mathcal{H}\,,\qquad
    \{S_a,L_1\}=-2\epsilon_{ab}Q^{\mathcal{Y}}_b\mathcal{H}\,,
\end{equation}
\begin{equation}\label{tK2QY2}
     [Q^{\mathcal{Y}}_a,S_b]=2i\delta_{ab}L_2\,,\qquad
    [Q^{\mathcal{Y}}_1,Q^{\mathcal{Y}}_2]=-2i\sigma_3
    P_{Q^{\mathcal{Y}}}(\mathcal{H})\,,\qquad
    [L_2,Q^{\mathcal{Y}}_a]=2iS_a
    P_{Q^{\mathcal{Y}}}(\mathcal{H})\,.
\end{equation}
\vskip0.2cm

All the integrals  (\ref{RTint}) including $\sigma_3$ but
excluding $\mathcal{RT}$ may be related between themselves
by unitary transformations, whose generators are
constructed in terms of the grading operators themselves.
For instance,
$U\sigma_3U^\dagger=\mathcal{R}\sigma_1=\tilde{\sigma_3}$,
$
    U=U^\dagger=U^{-1}=\frac{1}{\sqrt{2}}(\sigma_3+
    \mathcal{R}\sigma_1).
$ Being constructed from the integrals of motion, such a
transformation does not change the supersymmetric
Hamiltonian $\mathcal{H}$. On the other hand, if we apply
it to any nontrivial integral, the transformed operator
still will be an integral. Particularly, its application to
the integrals $S_1$ and $Q_1$ gives
\begin{equation}\label{SQtilde}
    \tilde{S}=i\mathcal{R}\sigma_2S_1={\rm diag}\,\left(
  \mathcal{RD}(x;\tau),
-\mathcal{RD}^\dagger(x;\tau) \right),\quad
    \tilde{Q}=i\mathcal{R}\sigma_2Q_1={\rm diag}\,\left(
  \mathcal{RA}(x;\tau),-\mathcal{RA}^\dagger(x;\tau)
\right).
\end{equation}
These are nontrivial hermitian \emph{nonlocal} integrals of
motion for the self-isospectral system
(\ref{Hcal})~\footnote{Notice that the (1+1)-dimensional GN
model has a system of infinitely many (nonlocal)
conservation laws.}. Eq. (\ref{SQtilde}) has a sense of
Foldy-Wouthuysen transformation that diagonalizes the
supercharges $S_1$ and $Q_1$. The price we pay for this is
a non-locality of the transformed operators.

Multiplication of (\ref{SQtilde}) by the grading operators
gives further nonlocal integrals, particularly,
$\sigma_3\tilde{S}$ and $\sigma_3\tilde{Q}$. Since the both
operators (\ref{SQtilde}) are diagonal, the Lam\'e
subsystem $H(x_+)$ may be characterized,  in addition to
the Lax integral $\mathcal{P}(x_+)$, by two nontrivial
nonlocal integrals
\begin{equation}\label{SQhat}
    \hat{S}=\mathcal{RD}(x;\tau)\,,\qquad
    \hat{Q}=\mathcal{RA}(x;\tau)\,.
\end{equation}
In correspondence with relations
$\mathcal{D}^\dagger(x;\tau)=-\mathcal{D}(x;-\tau)$ and
$\mathcal{A}^\dagger(x;\tau)=\mathcal{A}(x;-\tau)$, another
subsystem $H(x_-)$ is characterized then by the integrals
of the same form but with $\tau$ changed for $-\tau$. The
operator $\hat{\Gamma}=\mathcal{RT}$ is an integral for the
subsystem $H(x_+)$ [as well as for subsystem $H(x_-)$]. It
can be identified as a $\Z_2$ grading operator  that
assigns definite $\Z_2$ parities for nontrivial integrals
of the subsystem $H(x_+)$. Namely, in correspondence with
(\ref{RDAP1}) and (\ref{RDAP2}), the integrals
$-i\mathcal{P}(x_+)$ and $\hat{S}$ are fermionic operators
with respect to such a grading, while $\hat{Q}$ should be
treated as a bosonic operator. Multiplying fermionic
integrals by $i\hat{\Gamma}$ and bosonic integral by
$\hat{\Gamma}$, we obtain three more integrals for
$H(x_+)$. It is not difficult to calculate the
corresponding superalgebra generated by these integrals.
Let us note only that since the described supersymmetry may
be revealed in the subsystem $H(x_+)$ (or, in $H(x_-)$), it
may be treated as a bosonized supersymmetry, see
\cite{Hidsusy,CP1,CP2}. \vskip0.05cm

Let us return to the question of degeneration in our
self-isospectral system. This will allow us to observe some
other interesting properties related to the nonlocal
integrals (\ref{RTint}). Let us take a pair of mutually
commuting integrals  $S_1$ and $L_1$. They can be
simultaneously diagonalized, and for their common
eigenstates  we have $
    S_1\Psi^\alpha_{\pm,S_1,\epsilon}=
    \epsilon\mathcal{M}^{\mathcal{D}}(\alpha,\tau)
    \Psi^\alpha_{\pm,S_1,\epsilon}
$ and $ L_1\Psi^\alpha_{\pm,S_1,\epsilon}=
    \mp\eta(\alpha)\sqrt{P(\alpha)}\,
    \Psi^\alpha_{\pm,S_1,\epsilon}$, see Eqs.
(\ref{S1eigen}) and (\ref{L1eigen}).  We can distinguish
all the four states by these relations for any value of the
energy within the valence and conduction bands, and each
two doublet states for the edges $E=0,k'{}^2,1$ of the
allowed bands when $\tau\neq (\frac{1}{2}+n){\rm {\bf K}}$.
However, in the case of $\tau=(\frac{1}{2}+n){\rm {\bf
K}}$, the two ground states of zero energy are annihilated
by the both operators $S_1$ and $L_1$, and cannot be
distinguished by them. In this special case the operator
$\sigma_3$ commutes with $S_1$ and $L_1$ on the subspace
$E=0$, and may be used to distinguish the two ground
states. It is necessary to remember, however, that
$\sigma_3$ does not commute with $S_1$ on the subspaces of
nonzero energy.

There is yet another possibility. According to Table
\ref{T1}, the local integrals $S_1$ and $L_1$ commute with
the nonlocal integral $\mathcal{T}\sigma_2$. We find then
\begin{equation}\label{Ts2Psi}
    \mathcal{T}\sigma_2\Psi^\alpha_{\pm,S_1,\epsilon}=
    i\epsilon e^{\mp i\varphi^{\mathcal{D}}(\alpha,\tau)}
    \Psi^\alpha_{\pm,S_1,\epsilon}\,,
\end{equation}
where we used relation (\ref{phaseexplicit}). The operator
$\mathcal{T}\sigma_2$ detects therefore the phase in the
structure of the eigenstates of $S_1$. By comparing two
supersymmetric systems with the shift parameters $\tau$ and
$\tau+{\rm {\bf K}}$, and by taking into account the $2{\rm
{\bf K}}$-periodicity of the $\Theta$ function in
(\ref{FD}) and the $2{\rm {\bf K}}$-anti-periodicity of
$\sn\,u$, we get from (\ref{phaseexplicit}) that
$e^{i\left(\varphi^{D}(\alpha,\tau+{\rm {\bf
K}})-\varphi^{D}(\alpha,\tau)\right)} =e^{\frac{i}{{\rm
{\bf K}}}\kappa(\alpha)\tau}$. Hence the integral
$\mathcal{T}\sigma_2$ makes, particularly, the same job as
a \emph{translation for the period} operator (which is also
a nonlocal integral for the system)\,: it allows us to
determine an energy-dependent quasi-momentum. Finally, in
the case of zero energy ($\alpha={\rm {\bf K}}+i{\rm {\bf
K}}'$), treating $\tau=(\frac{1}{2}+n){\rm {\bf K}}$ as a
limit case,  one can also distinguish two ground states in
the supersymmetric doublet by means of (\ref{Ts2Psi}).

 Instead of $S_1$, $L_1$ and
$\mathcal{T}\sigma_2$, we could choose the triplet $S_2$,
$L_1$ and $\mathcal{T}\sigma_1$ of mutually commuting
integrals, see Table \ref{T1}. The states within the
supermultiplets can be distinguished also by choosing the
triplets of mutually commuting integrals ($Q_1$, $L_1$,
$\mathcal{T}\sigma_1$), or ($Q_2$, $L_1$,
$\mathcal{T}\sigma_2$). For the two latter cases, the
doublet of the ground states is annihilated by $Q_a$ and
$L_1$ for any value of the shift parameter $\tau$
(excluding the case $\tau=(\frac{1}{2}+n){\rm {\bf K}}$
when $Q_a$ are not defined), but the corresponding
integrals $\mathcal{T}\sigma_1$ or $\mathcal{T}\sigma_2$ do
here the necessary job of distinguishing the states as
well.

The integrals $\mathcal{R}\sigma_1$ and
$\mathcal{RT}\sigma_3$ act on the eigenstates of $S_1$,
with which they also commute, as $
    \mathcal{R}\sigma_1\Psi^\alpha_{\pm,S_1,\epsilon}(x,\tau)=
    -\epsilon e^{\pm i\varphi^{\mathcal{D}}(\alpha,\tau)}
    \Psi^\alpha_{\mp,S_1,\epsilon}(x,\tau), $
$
\mathcal{RT}\sigma_3\Psi^\alpha_{\pm,S_1,\epsilon}(x,\tau)=
    -
    \Psi^\alpha_{\mp,S_1,\epsilon}(x,\tau).$
These operators interchange the states with $+$ and $-$
indexes, and anti-commute with the integral $L_1$. The edge
states, which do not carry such an index,  are annihilated
by $L_1$, so that there is no contradiction with the
information presented in Table \ref{T1}.

\vskip0.05cm In conclusion of this section we note that the
Witten index computed with the grading operator identified
with any of the six nonlocal integrals (\ref{RTint}) is the
same as for a choice $\Gamma=\sigma_3$, i.~e. $\Delta_W=0$.

\section{Supersymmetry of the associated  periodic BdG
system}\label{susyBdGsec}

Till the moment we have discussed the self-isospectrality
of the one-gap Lam\'e system with the second order
Hamiltonian. Though we have shown that its supersymmetric
structure is much more rich than a usual one, from the
viewpoint of the physics of the GN model it is more natural
to look at the revealed picture from another perspective.

Let us take one of the first order integrals $S_a$ of the
self-isospectral Lam\'e system, say $S_1$, and consider it
as a first order, Dirac Hamiltonian. In such a way we
obtain an intimately related, but different physical
system. Unlike the second order operator $\mathcal{H}$, the
spectrum (\ref{S1eigen}) of $S_1$ depends on  $\tau$. We
get a periodic Bogoliubov-de Gennes system with Hamiltonian
$H_{BdG}=S_1$. The interpretation of the function
$\Delta(x;\tau)$ changes in this case\,: this is a Dirac
scalar potential in correspondence with a discussion from
section \ref{introd}. In dependence on a physical context,
it takes a sense of an order parameter, a condensate, or a
gap function.

\begin{figure}[h!]\begin{center}
\includegraphics[scale=0.6]{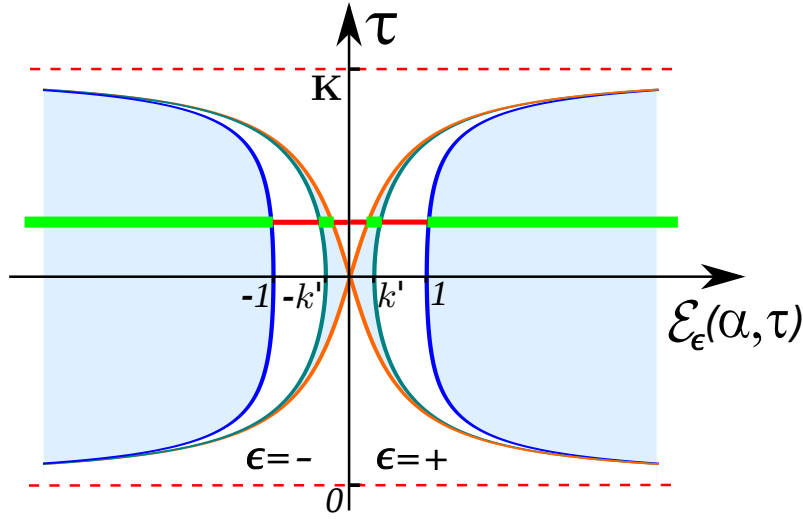}
\caption{Spectrum of $H_{BdG}=S_1$ possesses symmetries
$\mathcal{E}_\epsilon(\alpha,\tau)=
\mathcal{E}_\epsilon(\alpha,-\tau)=
\mathcal{E}_\epsilon(\alpha,\tau+{\rm {\bf K}})$,
$\mathcal{E}_\epsilon(\alpha,\frac{1}{2}{\rm {\bf
K}}+\tau)=\mathcal{E}_\epsilon(\alpha,\frac{1}{2}{\rm {\bf
K}}-\tau)$, and
$\mathcal{E}_-(\alpha,\tau)=-\mathcal{E}_+(\alpha,\tau)$.
Horizontal line shows a spectrum  for some value of $\tau$,
$\frac{1}{2}{\rm {\bf K}}<\tau<{\rm {\bf K}}$. The allowed
(forbidden) bands on it are presented by thick green (thin
red) intervals, whose points are distinguished by the
parameter $\alpha$, see Eq. (\ref{BdGvarE}). Curves
indicate the edges of the allowed bands (\ref{BdGvaledge}).
The point $\mathcal{E}_\epsilon({\rm {\bf K}}+i{\rm {\bf
K}}', \frac{1}{2}{\rm {\bf K}})=0$ corresponds to a doubly
degenerate energy level in the allowed band $[-k',k']$,
that is formed by the two merging at $\tau=\frac{1}{2}{\rm
{\bf K}}$ internal allowed  bands.}\label{fig3}
\end{center}
\end{figure}

The $\tau$-dependent spectrum of such a BdG system consists
of four or three allowed bands located symmetrically with
respect to the level $\mathcal{E}=0$, see Figure
\ref{fig3}. Interpretation of the bands also changes and
depends on the physical context.
 For $\tau\neq (\frac{1}{2}+n){\rm {\bf
K}}$, the positive and negative `internal' bands are
separated by a nonzero gap
$\Delta\mathcal{E}(\tau)=2\sqrt{\varepsilon(\tau)}=
2|\cn\,2\tau\,\ns\,2\tau|$, which disappears at $\tau=
(\frac{1}{2}+n){\rm {\bf K}}$. The total number of gaps in
the spectrum is three in the case $\tau\neq
(\frac{1}{2}+n){\rm {\bf K}}$, $\mathcal{E}\in
(-\infty,\mathcal{E}_{3,-}]\,\cup\,[\mathcal{E}_{2,-},
\mathcal{E}_{1,-}]\,\cup\,[\mathcal{E}_{1,+},\mathcal{E}_{2,+}]
\,\cup\, [\mathcal{E}_{3,+},\infty)$,
 while for $\tau=
(\frac{1}{2}+n){\rm {\bf K}}$ there are only two gaps,
$\mathcal{E}\in
(-\infty,\mathcal{E}_{3,-}]\,\cup\,[\mathcal{E}_{2,-},
\mathcal{E}_{2,+}]\,\cup\, [\mathcal{E}_{3,+},\infty)$.
According to (\ref{psiiCi}), (\ref{Medge}) and
(\ref{S1eigen}), the edges $\mathcal{E}_{i,\epsilon}$ of
the internal ($i=1,2$) and external ($i=3$) allowed bands
are
\begin{equation}\label{BdGvaledge}
    \mathcal{E}_{1,\epsilon}(\tau)=\epsilon
    \sqrt{\varepsilon(\tau)}\,,\qquad
    \mathcal{E}_{2,\epsilon}(\tau)=\epsilon
    \sqrt{k'{}^2+\varepsilon(\tau)}\,,\qquad
    \mathcal{E}_{3,\epsilon}(\tau)=\epsilon
    \sqrt{1+\varepsilon(\tau)}\,,
\end{equation}
where $\epsilon=\pm$, and the eigenstates have a form $
    \Psi_{i,\epsilon}(x,\tau)=\left(
   \psi_i(x_+),
  \epsilon e^{i\varphi^{\mathcal{D}}_i(\tau)}\psi_i(x_-)
\right)^T,$ $S_1\Psi_{i,\epsilon}(x,\tau)=
    \mathcal{E}_{i,\epsilon}\Psi_{i,\epsilon}(x,\tau).$

In the context of the physics of conducting polymers, for
example, the internal bands are referred to as the lower,
$[\mathcal{E}_{2,-},\mathcal{E}_{1,-}]$, and upper,
$[\mathcal{E}_{1,+},\mathcal{E}_{2,+}]$, polaron bands; the
upper external band, $[\mathcal{E}_{3,+},\infty)$, is
called the conduction band;  the lower external band,
$(-\infty,\mathcal{E}_{3,-}]$, is referred to as the
valence band  \cite{SaxBish}. In general case for
eigenstates (\ref{S1eigen}) we have
\begin{equation}\label{BdGvarE}
    S_1\Psi^\alpha_{\pm,S_1,\epsilon}(x,\tau)=
    \mathcal{E}_\epsilon(\alpha,\tau)
    \Psi^\alpha_{\pm,S_1,\epsilon}(x,\tau)\,,\qquad
    \mathcal{E}_\epsilon(\alpha,\tau)=\epsilon
    \sqrt{E(\alpha)+\varepsilon(\tau)}\,,
\end{equation}
where $E(\alpha)$  for internal  and external  bands is
given by Eqs. (\ref{E1band1}) and (\ref{E1band2}).

Since $H_{BdG}=S_1$ does not distinguish index $\pm$ of the
wave functions within the allowed bands, each corresponding
energy level is doubly degenerate. Six edge states for
$\tau\neq (\frac{1}{2}+n){\rm {\bf K}}$ are singlets. In
the case $\tau= (\frac{1}{2}+n){\rm {\bf K}}$, four edge
states with energies $\mathcal{E}=\pm k'$ and $\pm 1$ are
singlets. Zero energy states $\Psi_{1,\epsilon}$ form a
doublet in this case, as it happens for any other energy
level inside any allowed band.

 The described  degeneration in the spectrum of $S_1$
indicates that the BdG system might possess its \emph{own
nonlinear supersymmetric structure}. This is indeed the
case. First of all, from Table \ref{T1} we see that there
are three operators, $\mathcal{R}\sigma_1$,
$\mathcal{T}\sigma_2$ and $\mathcal{RT}\sigma_3$, which
commute with $S_1$, and square of each equals one. Hence,
each of them may be chosen  as a $\Z_2$ grading operator
for the BdG system. There are three more, nontrivial local
integrals of motion for $H_{BdG}$. One is the second order
operator $\mathcal{H}$. This, however, is not interesting
from the viewpoint of a supersymmetric structure since it
is just a shifted square of $H_{BdG}=S_1$,
$\mathcal{H}=S_1^2-\varepsilon(\tau)$. Then we have a third
order integral $L_1\equiv \mathcal{L}_1$, that has been
identified before as the Lax operator for the
self-isospectral Lam\'e system $\mathcal{H}$. Finally, the
fourth order operator $\mathcal{G}_1=S_1\mathcal{L}_1$ is
also identified as a local integral of motion. Note that
the both integrals $\mathcal{L}_1$ and $\mathcal{G}_1$
distinguish the states inside the allowed bands, which
differ in index $\pm\,$. On distinguishing the states with
$\mathcal{E}=0$ to be present in the spectrum if $\tau=
(\frac{1}{2}+n){\rm {\bf K}}$, see a discussion at the end
of the previous section. Further nontrivial, but nonlocal
integrals may be obtained if we multiply local integrals by
the operators $\mathcal{R}\sigma_1$, $\mathcal{T}\sigma_2$
and $\mathcal{RT}\sigma_3$. Then, as in the case of the
self-isospectral Lam\'e system, different choices for the
grading operator lead to distinct identifications of $\Z_2$
parities of the integrals.

For the sake of definiteness, let us choose
$\Gamma=\mathcal{R}\sigma_1$, and assume first that $\tau
\neq (\frac{1}{2}+n){\rm {\bf K}}$. Other two possibilities
for the choice of $\Gamma$ may be considered in an
analogous way. If, additionally, we restrict our analysis
by the integrals that do not include in their structure a
nonlocal in $\tau$ operator $\mathcal{T}$, we get two
$\Z_2$-even (commuting with $\Gamma$) integrals in addition
to $H_{BdG}=S_1$, namely, $\mathcal{R}\sigma_1$ and
$\mathcal{R}\sigma_1 S_1$. The four $\Z_2$-odd
(anticommuting with $\Gamma$) integrals are
$\mathcal{L}_1$, $\mathcal{G}_1$,
$\mathcal{L}_2=i\mathcal{R}\sigma_1\mathcal{L}_1$ and
$\mathcal{G}_2=i\mathcal{R}\sigma_1 \mathcal{G}_1$. All
these integrals are hermitian operators. It is interesting
to note that a nonlocal integral $\mathcal{R}\sigma_1 S_1$
is related to one of the diagonal nonlocal operators from
(\ref{SQtilde}),  $\mathcal{R}\sigma_1
S_1=\sigma_3\tilde{S}$. A  nonlocal diagonal operator
$\mathcal{G}_2$ also may be related to (\ref{SQtilde}),
$\mathcal{G}_2=\tilde{Q}S_1^2+
\mathcal{C}(\tau)\tilde{S}(S_1^2-\varepsilon(\tau))$.
Since, however, integrals $\mathcal{R}\sigma_1 S_1$ and
$\mathcal{G}_a$ are just the integrals
$\mathcal{R}\sigma_1$ and $\mathcal{L}_a$ multiplied by the
BdG Hamiltonian $S_1$, we can forget them as well as
$\mathcal{H}$. We obtain then nontrivial (anti)commutation
relations of the nonlinear BdG superalgebra,
\begin{equation}\label{BdGsusy1}
    [\mathcal{R}\sigma_1,\mathcal{L}_a]=-2i
    \epsilon_{ab}\mathcal{L}_b\,,\qquad
    \{\mathcal{L}_a,\mathcal{L}_b\}=2\delta_{ab}\hat{P}(S_1,\tau)\,.
\end{equation}
Here, in correspondence with Eqs. (\ref{PP}), (\ref{SSQQ})
and (\ref{RTs2}), $\hat{P}(S_1,\tau)$ is the six order
spectral polynomial of the BdG system,
\begin{equation}\label{BdGspecpoly}
    \hat{P}(S_1,\tau)=
    (S_1^2-\varepsilon(\tau))(S_1^2-\varepsilon(\tau)-k'{}^2)
(S_1^2-\varepsilon(\tau)-1)\,,
\end{equation}
whose six roots correspond to the energy levels
(\ref{BdGvaledge}).

Superalgebra (\ref{BdGsusy1}) has a structure similar to
that of a hidden, bosonized supersymmetry \cite{Hidsusy} of
the unextended Lam\'e system (\ref{HLame}), which was
revealed in \cite{CP2}.  There, the role of the grading
operator is played by a reflection operator $\mathcal{R}$,
the matrix integrals $\mathcal{L}_a$ are substituted  by
the Lax operator $-i\mathcal{P}(x)$, see Eq. (\ref{Zx+t}),
and by $\mathcal{R}\mathcal{P}(x)$. The six order
polynomial $\hat{P}(S_1,\tau)$ of the BdG Hamiltonian $S_1$
is changed there for a third order spectral polynomial
$P(H)$, see Eq. (\ref{PP}). \vskip0.05cm

We have seen that the structure of the BdG spectrum changes
significantly at $\tau =(\frac{1}{2}+n){\rm {\bf K}}$.
Essential  changes happen also in the superalgebraic
structure. {}Indeed, from (\ref{AQYeps}) it follows that
$[S_1,Q^{\mathcal{Y}}_2]=2i
\sigma_3\epsilon_{ab}\mathcal{C}(\tau) \varepsilon(\tau)$,
i.~e. in a generic case $Q^{\mathcal{Y}}_2$ does not
commute with $H_{BdG}$. In contrary, for $\tau
=(\frac{1}{2}+n){\rm {\bf K}}$ this is an additional
nontrivial, second order integral of motion  of the BdG
system. This integral, like the third order integral $L_1$,
also distinguishes the states marked by the index $\pm$
inside the allowed bands,
$Q^{\mathcal{Y}}_2\Psi^\alpha_{\pm,S_1,\epsilon}= \pm\eta
\sqrt{P_{Q^{\mathcal{Y}}}(E(\alpha))}\,\Psi^\alpha_{\pm,S_1,\epsilon}$,
where $\eta$ is the same as in (\ref{dkdE}) and
(\ref{LaxBloch}), i.e. $\eta=-1$ for $0\leq E\leq k'{}^2$
and $\eta=+1$ for $E\geq 1$, while $P_{Q^{\mathcal{Y}}}(E)$
is a polynomial that appeared earlier in (\ref{PQYH}),
i.~e. $P_{Q^{\mathcal{Y}}}(E)=(E-k'{}^2)(E-1)$. In this
case, $L_1$ is not independent integral for the BdG system
anymore since here $L_1=-S_1Q^{\mathcal{Y}}_2$ in
correspondence with (\ref{S*Q=L}).  Integral
$Q^{\mathcal{Y}}_2$ anticommutes with $\mathcal{R}\sigma_1$
and $\mathcal{RT}\sigma_3$. Let us choose, again,
$\Gamma=\mathcal{R}\sigma_1$, and denote
$\mathcal{Q}_1=Q^{\mathcal{Y}}_2$ and
$\mathcal{Q}_2=i\Gamma\mathcal{Q}_1$.  Instead of
(\ref{BdGsusy1}),  we get  a nonlinear superalgebra of the
order four,
\begin{equation}\label{BdGsusy2}
    [\mathcal{R}\sigma_1,\mathcal{Q}_a]=-2i
    \epsilon_{ab}\mathcal{Q}_b\,,\qquad
    \{\mathcal{Q}_a,\mathcal{Q}_b\}=2\delta_{ab}\hat{P}_{\mathcal{Q}}(S_1)\,,
\end{equation}
where $\hat{P}_{\mathcal{Q}}(S_1)=(S_1^2-k'{}^2)(S_1^2-1)$.
\vskip0.2cm

It is interesting to see what happens with the Witten index
in the described unusual supersymmetry of the BdG system
with the first order Hamiltonian. One can construct the
eigenstates of the grading operator
$\Gamma=\mathcal{R}\sigma_1$,
\begin{equation}\label{Gammaeigen}
    \Gamma \Psi^{(\epsilon)}(x;\alpha,\tau)=
    -\epsilon \Psi^{(\epsilon)}(x;\alpha,\tau)\,,\quad
    \Psi^{(\epsilon)}(x;\alpha,\tau)\equiv
    \Psi^\alpha_{+,S_1,\epsilon}(x,\tau)
    +e^{i\varphi^{\mathcal{D}}(\alpha,\tau)}
    \Psi^\alpha_{-,S_1,\epsilon}(x,\tau)\,.
\end{equation}
For any energy value inside any allowed band (including
$\mathcal{E}=0$ in the case of $\tau= (\frac{1}{2}+n){\rm
{\bf K}}$), we have two states with opposite eigenvalues of
$\Gamma$, and these contribute zero into the Witten index
$\Delta_W={\rm Tr}\,  \Gamma$, where trace is taken over
all the eigenstates of the grading operator $\Gamma$. On
the other hand, the edge states $\Psi_{i,\epsilon}(x,\tau)$
are singlets. They are also the eigenstates of $\Gamma$.
The eigenstates of opposite energy signs have opposite
eigenvalues, $+1$ and $-1$, of the grading operator. As a
result, we conclude that the Witten index $\Delta_W$ in
such a supersymmetric system equals zero for any value of
$\tau$ [i.e., for $\tau \neq (\frac{1}{2}+n){\rm {\bf K}}$
when there are no zero energy states in the spectrum, and
for $\tau = (\frac{1}{2}+n){\rm {\bf K}}$ when the spectrum
contains a doublet of zero energy states], like this
happens in the self-isospectral Lam\'e system with the
second order supersymmetric Hamiltonian. The same result
$\Delta_W=0$ is obtained for the choices
$\Gamma=\mathcal{T}\sigma_2$ and
$\Gamma=\mathcal{RT}\sigma_3$.

Finally, it is worth to notice that in accordance with the
structure of superalgebra (\ref{BdGsusy1}), the third order
matrix BdG supercharges $\mathcal{L}_a$ annihilate all the
six edge eigenstates of $H_{BdG}=S_1$ in the case of
$\tau\neq (\frac{1}{2}+n){\rm {\bf K}}$. In special cases
$\tau= (\frac{1}{2}+n){\rm {\bf K}}$ a central gap
disappears in the spectrum, and, consistently  with
(\ref{BdGsusy2}), all the remaining four edge states are
the zero modes of the second order matrix BdG supercharges
$\mathcal{Q}_a$. In other words, the spectral changes that
happen in the BdG system at special values of the parameter
$\tau= (\frac{1}{2}+n){\rm {\bf K}}$, which correspond to a
zero value of the bare mass $m_0$ in the GN model
(\ref{LagGN}), are reflected coherently by the changes in
its superalgebraic structure.

\section{Infinite period limit}\label{non-period}

Let us discuss now the infinite period  limit of our
self-isospectral Lam\'e and the associated BdG systems, i.
e. the case when the period $2{\rm {\bf K}}$  tends to
infinity.

${\rm {\bf K}}\rightarrow \infty$  assumes~\footnote{Any of
these four limits assumes three others.} $k\rightarrow 1$,
$k'\rightarrow 0$, ${\rm {\bf K}}'\rightarrow
\frac{1}{2}\pi$, and relations (\ref{limit2}), and
(\ref{Z2}) have to be employed. According to (\ref{Z2}) and
(\ref{Theta}), a limit for  a quotient of $\Theta$
functions is also well defined,
\begin{equation}\label{Thetalimit}
    \lim_{k\rightarrow
    1}\frac{\Theta(u)}{\Theta(v)}=
    \frac{\cosh(u)}{\cosh(v)}\,,\qquad u,\, v\, \in \C\,.
\end{equation}

Periodic Lam\'e Hamiltonian (\ref{HLame}) transforms in
this limit into a reflectionless one-gap P\"oschl-Teller
Hamiltonian
\begin{equation}\label{HPT}
    H_{PT}(x)=-\frac{d^2}{dx^2}-\frac{2}{\cosh^2 x}+1\,.
\end{equation}
When the limit ${\rm {\bf K}}\rightarrow\infty$ is applied
to the self-isospectral system (\ref{Hcal}), we assume that
a shift parameter $\tau$ remains to be finite. As a result
we get a self-isospectral non-periodic PT system,
\begin{equation}\label{HPTself}
    \mathcal{H}_{PT}(x)={diag}\,
    (H_\tau(x),H_{-\tau}(x))\,,
\end{equation}
where $H_\tau(x)=H_{PT}(x+\tau)$ and
$H_{-\tau}(x)=H_{PT}(x-\tau)$. In what follows we trace out
how the peculiar supersymmetry of the self-isospectral
Lam\'e system transforms in the infinite period limit into
the supersymmetric structure of the system (\ref{HPTself}),
which was studied recently in \cite{PRDPT}.

Since the super-partners in (\ref{HPTself}) are the two
mutually shifted copies of the same PT system, it is clear
that the limit does not change the Witten index\,: it
remains to be equal zero as in the periodic case. In
general, however, the index may or may not change depending
on the concrete form of the self-isospectral Lam\'e system
to which the limit is applied. For instance, in the case of
the system with superpartners $H(x)$ and $H(x+{\rm {\bf
K}})$ [see a remark just below Eq. (\ref{Hx+t})], the
infinite-period limit gives, instead of (\ref{HPTself}), a
supersymmetric system with one superpartner to be the PT
system (\ref{HPT}), while another one (which is a  limit of
$H(x+{\rm {\bf K}})$) to be a free particle
$H_0=-\frac{d^2}{dx^2}+1$. Superpartner potentials in such
a supersymmetric (but not self-isospectral) system are
distinct. The only difference of the spectrum for the
system (\ref{HPT}) from that for $H_0$ consists in the
presence of a unique bound state, see below. Consequently,
the Witten index changes in the infinite period limit, by
taking a value of the modulus one. If in the system
(\ref{Hcal}) one takes $\tau=\tau({\rm {\bf K}})$ such that
$\tau\rightarrow \infty$ for ${\rm {\bf K}}\rightarrow
\infty$, the limit produces then a trivial self-isospectral
system composed from the two copies of the free particle
Hamiltonian $H_0$. In such a case, the Witten index does
not change in agreement with (\ref{HPTself}) and
(\ref{HPT}).

The listed examples also mean that the shifts for the
period, in a sense, `interfere' with  the infinite period
limit. Self-isospectral Lam\'e system composed from
$H(x_+)$ and $H(x_-)$ is equivalent, for instance, to a
system with super-partner Hamiltonians $H(x_+)$ and
$H(x_-+2{\rm {\bf K}})$~\footnote{The second system,
however, is characterized by another phase
(\ref{phaseexplicit}) with $\tau$ changed for $\tau-{\rm
{\bf K}}$.}. If before taking a limit we do not `eliminate'
the period $2{\rm {\bf K}}$ shift in the second subsystem,
we will obtain a (not self-isospectral) system with
super-partners $H_\tau$ and $H_0$ instead of
(\ref{HPTself}).

Let us return to the symmetric case of the self-isospectral
Lam\'e system (\ref{Hcal}), whose infinite period limit
corresponds to the self-isospectral PT system
(\ref{HPTself}). All the energy values (\ref{E1band1}) of
the valence band transform  into zero in the infinite
period limit because of  $k'\rightarrow 0$, i. e. all this
band shrinks just into a one energy level $E=0$ for the
system (\ref{HPT}). In conformity with this, all the Bloch
states (\ref{Psipm}) of this band, including the edge
states $\dn\,x$ and $\cn\,x$, turn into a unique bound
state $\frac{1}{\cosh x}$ of $E=0$ for PT
system~\footnote{The states (\ref{Psipm}) for the valence
band should be `renormalized' (divided) by a constant
$\Theta({\rm {\bf K}})/\Theta(0)$ to cancel the
multiplicative factor that diverges in the limit ${\rm {\bf
K}}\rightarrow\infty$ in correspondence with
(\ref{Thetalimit}).}. Then  the states $1/\cosh (x\pm
\tau)$ form a supersymmetric doublet of the ground states
for self-isospectral system (\ref{HPTself}). The doublet of
the edge states $\sn\,(x\pm \tau)$ of the system
(\ref{Hcal}) transforms into a doublet of the lowest states
$\tanh(x\pm\tau)$ of the energy $E=1$ in the scattering
sector of the spectrum for (\ref{HPTself}). It is
interesting to see how the eigenstates with $E>1$ in the
scattering sector of the PT system originate from the Bloch
states (\ref{Psipm}). The energy (\ref{E1band2}) as a
function of the parameter $\beta$, which in the infinite
period limit takes values in the interval $0\leq
\beta<\frac{\pi}{2}$, reduces to $
    E(i\beta)=\frac{1}{\cos^2\beta}\geq 1.
$ The states (\ref{Psipm}) transform into
$\Psi^{i\beta}_\pm(x)=\cos\beta\left(\tanh x\pm i\tan
\beta\right) \exp(\mp i x\tan\beta)$. Denoting $\tan
\beta={\Bbbk}\geq 0$,  we obtain $E=1+\Bbbk^2$, and the
states $\Psi^{i\beta}_\mp(x)$ take the form of the
scattering eigenstates of the PT system, $
    \Psi^{i\beta}_\mp(x)\longrightarrow\,\,\,
    \Psi^{\pm \Bbbk}(x)=-\frac{1}{\sqrt{E}}(\pm i\Bbbk-\tanh x)e^{\pm
    i\Bbbk x}.
$
\vskip0.05cm

We have
\begin{equation}\label{Flimit}
    F(x;\tau)\xrightarrow{k\rightarrow 1}\,
    \frac{\cosh x_-}{\cosh x_+}\,e^{x\coth
2\tau}
\end{equation}
for function (\ref{c02t}), cf. Eq. (5.17) in \cite{PRDPT}.
In correspondence with (\ref{Hve2t}), this is a nonphysical
eigenstate of $H_\tau$ of eigenvalue $-1/\sinh^2 2\tau$.
Function $\Delta(x;\tau)$ in the form (\ref{DelZeta})
transforms  into
\begin{equation}\label{Deltalim}
    \Delta(x;\tau)\xrightarrow{k\rightarrow 1}\, \Delta_\tau(x)=\coth
    2\tau +\tanh x_- -\tanh x_+\,,
\end{equation}
while Eq. (\ref{Del1}) gives, equivalently,
\begin{equation}\label{Deltalim2}
    \Delta(x;\tau)\xrightarrow{k\rightarrow 1}\, \Delta_\tau(x)=\frac{2}{\sinh
    4\tau}+\tanh 2\tau \tanh x_-\tanh x_+\,.
\end{equation}
Non-periodic superpotential (gap function) (\ref{Deltalim})
corresponds to the DHN kink-antikink baryons  \cite{DaHN}.
For the first order intertwining operator we have
\begin{equation}\label{DXlimit}
    \mathcal{D}(x;\tau)
      \xrightarrow{k\rightarrow 1}\frac{d}{dx}-\Delta_\tau(x)
    \equiv X_\tau\,,
\end{equation}
cf. (2.26) in \cite{PRDPT}. It is the operator that appears
in the limit structure of the supercharges $S_a$,
\begin{equation}\label{S1limit}
    S_1 \xrightarrow{k\rightarrow 1}\, \left(%
\begin{array}{cc}
  0 & X_\tau^\dagger \\
  X_\tau & 0 \\
\end{array}%
\right)\equiv S_{PT,1}\,,\qquad
    S_2\xrightarrow{k\rightarrow 1}\,S_{PT,2}=i\sigma_3
    S_{PT,1}\,.
\end{equation}
For the second order intertwining operator (\ref{Axt}),
\begin{equation}\label{Alimit}
    \mathcal{A}(x;\tau)\xrightarrow{k\rightarrow
1}\,A_{-\tau}A^\dagger_\tau\equiv Y_\tau\,,
\end{equation}
where
$
    \lim_{{\rm {\bf K}}\rightarrow\infty}
    \mathcal{D}(x+\tau+\frac{1}{2}{\rm {\bf K}};\frac{1}{2}{{\rm {\bf K}}})=
    \lim_{{\rm {\bf K}}\rightarrow\infty}
    \mathcal{D}(x+\frac{1}{2}{\rm {\bf K}};-\tau+\frac{1}{2}{\rm {\bf K}})=
    \frac{d}{dx}-\tanh x_+\equiv A_\tau(x),$
and $A_{-\tau}$ is obtained via the change $\tau\rightarrow
-\tau$. A limit of the second order integrals (\ref{Q}) is
\begin{equation}\label{QlimY}
    Q_1 \xrightarrow{k\rightarrow
    1}\,\left(%
\begin{array}{cc}
  0 & Y_\tau^\dagger \\
  Y_\tau & 0 \\
\end{array}%
\right)\equiv Q_{PT,1}\,,\qquad
    Q_2\xrightarrow{k\rightarrow 1}\,Q_{PT,2}=i\sigma_3
    Q_{PT,1}\,,
\end{equation}
cf. Eq. (2.18) in \cite{PRDPT}. The first order operators
$A_\tau$ and $A_{-\tau}$ factorize also the
self-isospectral pair of the PT Hamiltonians,
$H_\tau=A_\tau A_\tau^\dagger$,
$H_{-\tau}=A_{-\tau}A^\dagger_{-\tau}$, as well as a free
particle Hamiltonian, $H_0=A_\tau^\dagger A_\tau=
A_{-\tau}^\dagger A_{-\tau}$.

The phases that appear in the action of the intertwining
operators  $\mathcal{D}(x;\tau)$ and  $\mathcal{A}(x;\tau)$
on the super-partner's eigenstates, see Eqs.
(\ref{Dintersusy}) and (\ref{Aintersusy}), transform into
\begin{equation}\label{phiDAlim}
    e^{i\varphi^{\mathcal{D}}(\alpha,\tau)}\xrightarrow{k\rightarrow 1}\,
    e^{-2i\Bbbk\tau}\cdot \frac{-i\Bbbk -\coth 2\tau}
    {\sqrt{\Bbbk^2+\coth^22\tau}}\,,\qquad
    e^{i\varphi^{\mathcal{A}}(\alpha,\tau)}\xrightarrow{k\rightarrow 1}\,
    e^{-2i\Bbbk\tau}\,.
\end{equation}
They are associated with the action of the intertwining
operators $X_\tau$ and $Y_\tau$ on the eigenstates of
super-partner systems $H_\tau$ and $H_{-\tau}$, and appear
in the structure of the eigenstates of the first,
(\ref{S1limit}), and the second, (\ref{QlimY}), order
integrals of the self-isospectral PT system \cite{PRDPT}.

By employing a relation
$2\mathcal{P}(x_+)=\mathcal{D}^\dagger(x;\tau)\mathcal{A}(x;\tau)-
\mathcal{A}^\dagger(x;\tau)\mathcal{D}(x;\tau)$ that
follows from Eq. (\ref{DAAD}), we find that
\begin{equation}\label{PZlim}
    \mathcal{P}(x_+)\xrightarrow{k\rightarrow 1}\,
    A_\tau\frac{d}{dx}A_\tau^\dagger\equiv Z_\tau\,,
\end{equation}
cf. (2.24) in \cite{PRDPT}.  For the limit of the Lax
integrals we get then
\begin{equation}\label{L1limit}
    L_1\xrightarrow{k\rightarrow 1}\,
    -i\left(%
\begin{array}{cc}
  Z_\tau & 0 \\
  0 & Z_{-\tau} \\
\end{array}%
\right)\equiv L_{PT,1}\,,\qquad
    L_2\xrightarrow{k\rightarrow 1}\,L_{PT,2}=\sigma_3
    L_{PT,1}\,.
\end{equation}
Finally, for a constant
$\mathcal{C}(\tau)=\ns\,2\tau\,\nc\,2\tau\,\dn\,2\tau$ that
appears  in the superalgebraic (anti)commutation relations
of our system we obtain
\begin{equation}\label{Ctaulim}
    \mathcal{C}(\tau)\xrightarrow{k\rightarrow 1}\, \coth
    2\tau\equiv \mathcal{C}_{2\tau}\,,
\end{equation}
cf. the first term in Eq. (\ref{Deltalim}).

With the described infinite period limit relations, we find
a correspondence between the supersymmetric structures in
the self-isospectral one-gap Lam\'e and PT systems.
Particularly, applying the infinite period limit to the
superalgebraic relations of the self-isospectral Lam\'e
system and making use of the described correspondence, one
may reproduce  immediately the superalgebraic relations for
the self-isospectral PT system (\ref{HPTself}).

The same $\tau$-dependent constant
$\mathcal{C}_{2\tau}=\coth 2\tau$ shows up in
representation for superpotential (\ref{Deltalim}) and in
the superalgebraic structure for the self-isospectral
non-periodic PT system (\ref{HPTself}) due to relation
(\ref{Ctaulim}). Notice, however, that corresponding
functions of a shift parameter, ${\rm z}(\tau)$ and
$\mathcal{C}(\tau)$, which appear in the periodic system,
are different. In the next section we will return to this
observation. \vskip0.05cm

The infinite-period limit of the second order intertwining
operator $\mathcal{Y}(x;\tau)$ may be found by employing
relation (\ref{Axt}),
\begin{equation}\label{Ylimitx-t}
    \lim_{K\rightarrow\infty} \mathcal{Y}(x;\tau)=
    -Y_\tau-\mathcal{C}_{2\tau}X_\tau\,.
\end{equation}
 It plays no special
role in the supersymmetric structure of the
self-isospectral PT system (\ref{HPTself}). Let us,
however, shift $x\rightarrow x-\tau$ in (\ref{Ylimitx-t})
and then take a limit $\tau\rightarrow\infty$. Such a
double limit procedure applied to the self-isospectral
Lam\'e system $\mathcal{H}$ produces a non-periodic
supersymmetric system $\hat{\mathcal{H}}=diag\,(H_{PT}(x),
H_0(x))$ that is composed from the PT system (\ref{HPT})
and the free particle $H_0=-\frac{d^2}{dx^2}+1$. Operator
$\mathcal{Y}(x;\tau)$ in such a limit transforms into the
second order operator
$\hat{y}(x)=\frac{d}{dx}(\frac{d}{dx}+\tanh x)$ that
intertwines $H_{PT}$ with $H_0$,
$\hat{y}(x)H_{PT}(x)=H_0(x)\hat{y}(x)$. The kernel of
$\hat{y}$ is formed by singlet eigenstates $1/\cosh x$
($E=0$) and $\tanh x$ ($E=1$) of the PT system $H_{PT}(x)$,
cf. the discussion of the kernel of
$\mathcal{Y}(x;\frac{1}{2}{\rm {\bf K}})$ in section
\ref{highinter}. Hermitian conjugate operator
$\hat{y}^\dagger(x)$ intertwines as
$\hat{y}^\dagger(x)H_0(x)=H_{PT}(x)\hat{y}^\dagger(x)$, and
annihilates the eigenstate $1$ of the lowest energy $E=1$
and a non-physical state $\sinh x$ of zero energy in the
spectrum of $H_0$. Integrals $S_a$, $Q^{\mathcal{Y}}_a$ and
$L_a$  transform in such a double limit into the integrals
of the supersymmetric system $\hat{\mathcal{H}}$,
\begin{equation}\label{Ss1lim}
    S_1\rightarrow -\left(%
\begin{array}{cc}
  0 & A_0 \\
  A_0^\dagger & 0 \\
\end{array}%
\right)\equiv \hat{s}_1\,,\quad
    Q^{\mathcal{Y}}_1\rightarrow
    \left(%
\begin{array}{cc}
  0 & \hat{y}^\dagger \\
  \hat{y} & 0 \\
\end{array}%
\right)\equiv \hat{q}{}^{y}_1\,,\quad
    L_1\rightarrow
    -i\left(%
\begin{array}{cc}
  A_0\frac{d}{dx}A_0^\dagger & 0 \\
  0 & H_0\frac{d}{dx} \\
\end{array}%
\right)\equiv \hat{l}_1,
\end{equation}
and $S_2\rightarrow \hat{s}_2=i\sigma_3\hat{s}_1$,
$Q^{\mathcal{Y}}_2\rightarrow
\hat{q}^{y}_2=i\sigma_3\hat{q}^{y}_1$, $L_2\rightarrow
\hat{l}_2=\sigma_3\hat{l}_1$, where
$A_0=\lim_{\tau\rightarrow\infty}A_\tau(x-\tau)=\frac{d}{dx}-\tanh
x=A_0(x)$, and we have used the relations
$\lim_{\tau\rightarrow\infty}A_{-\tau}(x)=\frac{d}{dx}+1$,
and $A_0^\dagger A_0=H_0$, and
$\hat{y}=-\frac{d}{dx}A_0^\dagger$.

Non-periodic superpotential (gap function) $\Delta(x)=\tanh
x$ that appears in the structure of the first and second
order intertwining operators as well as in that of the
integrals (\ref{Ss1lim}) corresponds to the famous CCGZ
kink solution \cite{DaHN,PTDol,Gross} of the GN model.

From the total number of seven integrals of motion
(\ref{RTint}) and $\sigma_3$, each of which can be used as
a grading operator for self-isospectral Lam\'e and PT
systems, in the described double limit survive only
three\,: in addition to the obvious operator $\sigma_3$,
nonlocal operators $\mathcal{R}$ and $\mathcal{R}\sigma_3$
are also the integrals for supersymmetric system
$\hat{\mathcal{H}}$. The last two operators originate in
the double limit from the integrals $\mathcal{RT}$ and
$\mathcal{RT}\sigma_3$. Having in mind  this
correspondence, Table \ref{T1} still may be used for
identification of  $\Z_2$ parities of the integrals
$\hat{s}_a$, $\hat{q}^y_a$ and $\hat{l}_a$, and it is not
difficult to obtain corresponding forms for superalgebra
for each of the three possible choices of the grading
operator in this case, see  \cite{Tri,AdSPT}.

Let us look what happens here with the Witten index.  As we
discussed at the beginning of this section, the only
asymmetry between the spectra of the superpartner
Hamiltonians  $H_{PT}$ and $H_0$ is the presence of the
zero energy bound state in the first super-partner system, which is
described by the eigenstate $(1/\cosh x, 0)^T$ of
the supersymmetric system $\hat{\mathcal{H}}$.
The doublet with $E=1$ is formed by
the eigenstates $(\tanh x,0)^T$ and $(0,1)^T$.
The first state  is an eigenstate
of all the three operators $\sigma_3$,
$\mathcal{R}$ and $\mathcal{R}\sigma_3$ with the same
eigenvalue $+1$, while for the
 second and third states the eigenvalues are, respectively,
$+1$, $-1$, $-1$ and $-1$, $+1$, $-1$. All the forth-fold
degenerate energy levels in the scattering part of the
spectrum with $E>1$ contribute zero into the Witten index
$\Delta_W={\rm Tr}\, \Gamma$. As a result, for all the
three choices of the grading operator for non-periodic
supersymmetric system $\hat{\mathcal{H}}$ we have
consistently $|\Delta_W|=1$~\footnote{$\Delta_W$ takes
values $+1$ for $\Gamma=\sigma_3$ and $\mathcal{R}$, and
$-1$ for $\mathcal{R}\sigma_3$. A difference in sign is not
important, however, since it can be removed by changing a
sign in definition of the grading operator in the last
case.}.

On the other hand, the first order matrix operator
$\hat{s}_1$ is identified here as a  limit of the BdG
Hamiltonian $H_{BdG}=S_1$. As may be checked directly,
operator $\mathcal{R}\sigma_3$ commutes with $\hat{s}_1$ in
accordance with Table \ref{T1} if to take into account the
correspondence between nonlocal integrals discussed above.
Therefore, it can be identified as a grading operator for a
peculiar supersymmetry of the BdG system with the
Hamiltonian $\hat{h}_{BdG}=\hat{s}_1$, in which the second
order integral $\hat{q}^y_2$, and the nonlocal operator
$i\mathcal{R}\sigma_3 \hat{q}^y_2$ are identified as the
odd supercharges, and $\hat{l}_1=-\hat{s}_1\hat{q}^y_2$,
cf. (\ref{S*Q=L}). Corresponding superalgebra has a form
(\ref{BdGsusy2}) with obvious substitutions. The state
$(1/\cosh x, 0)^T$, is a unique zero mode of the first
order matrix hamiltonian $\hat{s}_1$, while  two states
$(\tanh x,\pm1)^T $ are the singlet eigenstates of
$\hat{s}_1$ of the eigenvalues $\pm 1$, which are also the
eigenstates of the grading operator $\mathcal{R}\sigma_3$
of the eigenvalue $-1$. \vskip0.1cm

Thus, the modulus of the Witten index changes from zero to
one for the supersymmetries of the both, second,
$\hat{\mathcal{H}}$, and first, $h_{BdG}=\hat{s}_1$, order
systems. This reflects effectively the changes in the
spectrum that happen in the described infinite-period limit
of the self-isospectral second order Lam\'e and the
associated first order BdG systems.

\section{Extended supersymmetric picture and Darboux dressing}

Let us discuss now another interesting aspect of our
self-isospectral periodic supersymmetric system in the
light of the infinite period limit. As it was shown in
\cite{PRDPT}, the supersymmetric structure of the
non-periodic self-isospectral system (\ref{HPTself}) has a
peculiar property\,: all its integrals can be treated as a
Darboux-dressed form of the integrals of a free particle
system $H_0(x)$. We clarify now what corresponds here, in
the periodic case, to the Darboux-dressing structure of the
self-isospectral PT system (\ref{HPTself}). For that, we
extend a picture related to the intertwining operators and
the Darboux displacements  associated with them.
\vskip0.05cm

Consider along with our self-isospectral supersymmetric
Lam\'e system (\ref{Hcal}),
$\mathcal{H}(x)=diag\,(H(x+\tau),H(x-\tau))$, its copy
shifted for the half period, $\mathcal{H}(x+{\rm {\bf
K}})=diag\,(H(x+{\rm {\bf K}}+\tau),H(x+{\rm {\bf
K}}-\tau)$. Any two of the four (single-component)
Hamiltonians may be connected  by intertwining relation of
the form
$\mathcal{D}(\xi;\mu)H(\xi+\mu)=H(\xi-\mu)\mathcal{D}(\xi;\mu)$.
Putting $\xi=x+\frac{1}{2}(\tau_1+\tau_2)$ and
$\mu=\frac{1}{2}(\tau_1-\tau_2)$, $\tau_1\neq \tau_2+2{\rm
{\bf K}}n$, we present this relation in a more appropriate
form
\begin{eqnarray}\label{tau1tau2}
    &\mathcal{D}(x+\frac{1}{2}(\tau_1+\tau_2);\frac{1}{2}(\tau_1-\tau_2))
    H(x+\tau_1)=
    H(x+\tau_2)\mathcal{D}(x+\frac{1}{2}(\tau_1+\tau_2);
    \frac{1}{2}(\tau_1-\tau_2))\,.\quad &
\end{eqnarray}
Here $\tau_1$ and $\tau_2$   take values in the set $\{
-\tau,\,\tau,\,-\tau+{\rm {\bf K}},\,\tau+{\rm {\bf K}}\}$,
and supersymmetric Hamiltonians $\mathcal{H}(x)$ and
$\mathcal{H}(x+{\rm {\bf K}})$ may be related by $
    \tilde{\mathcal{D}}\mathcal{H}(x+{\rm {\bf
    K}})=\mathcal{H}(x)\tilde{\mathcal{D}},
$
$
    \tilde{\mathcal{D}}^\dagger\mathcal{H}(x)=
    \mathcal{H}(x+{\rm {\bf
    K}})\tilde{\mathcal{D}}^\dagger,
$
where
\begin{eqnarray}\label{Ddiag}
    &\tilde{\mathcal{D}}={\rm diag}\,\left(
  \mathcal{D}(x+\tau+\frac{1}{2}{\rm {\bf
    K}};\frac{1}{2}{\rm {\bf
    K}}),
    \mathcal{D}(x-\tau+\frac{1}{2}{\rm {\bf
    K}};\frac{1}{2}{\rm {\bf
    K}})
\right).&
\end{eqnarray}

In general case, if any two Hamiltonians $h$ and
$\tilde{h}$ are related by intertwining operators $D$ and
$D^\dagger$, $Dh=\tilde{h}D$,
$hD^\dagger=D^\dagger\tilde{h}$, and if $J$ is an integral
for $h$, $[h,J]=0$, then the operator
$\tilde{J}=DJD^\dagger$ is an integral for $\tilde{h}$. The
system $\mathcal{H}(x)$ is characterized by the set of
local integrals of motion $J(x)=\{\sigma_3$, $S_a(x)$,
$Q_a(x)$, $L_a(x)\}$, while the system $\mathcal{H}(x+{\rm
{\bf K}})$, is described by the same but shifted set,
$J(x+{\rm {\bf K}})$. Identifying $\mathcal{H}(x+{\rm {\bf
K}})$, $\mathcal{H}(x)$ and $\tilde{\mathcal{D}}$ with $h$,
$\tilde{h}$ and $D$, respectively, we find that
$
    \tilde{J}=\tilde{\mathcal{D}}J(x+{\rm {\bf
    K}})\tilde{\mathcal{D}}^\dagger= J(x)\mathcal{H}(x).
$ In other words, the Darboux dressed integral of one
system is just the corresponding integral of another,
displaced self-isospectral periodic system, multiplied by
its Hamiltonian.  Nonlocal operators (\ref{RTint}), which
are the integrals for $\mathcal{H}(x)$, are also the
integrals of motion for the displaced system
$\mathcal{H}(x+{\rm {\bf K}})$. Then one finds that a
similar relation  is valid also for these nonlocal
integrals as well as for nontrivial diagonal nonlocal
integrals (\ref{SQtilde}). The only difference is that for
all the integrals that contain a factor $\mathcal{R}$,
including (\ref{SQtilde}),  there appears a minus sign,
like in $\tilde{\mathcal{D}}\tilde{S}(x+{\rm {\bf
    K}})\tilde{\mathcal{D}}^\dagger=-
\tilde{S}(x)\mathcal{H}(x)$. Notice also that the Darboux
dressed form of the trivial integral $\mathds{1}$ (that is
a unit two-by-two matrix) for the displaced system
$\mathcal{H}(x+{\rm {\bf K}})$ coincides with the
Hamiltonian $\mathcal{H}(x)$, $\tilde{\mathcal{D}}
\mathds{1} \tilde{\mathcal{D}}^\dagger = \mathcal{H}(x)$.

Since the both self-isospectral supersymmetric systems are
just two copies of the same periodic system shifted
mutually in the half period, the described picture is not
so unexpected.  Let us look, however, at this result from
another viewpoint. In the infinite period limit,
supersymmetric systems $\mathcal{H}(x)$ and
$\mathcal{H}(x+{\rm {\bf K}})$ transform, respectively,
into (\ref{HPTself}) and
\begin{equation}\label{calH0}
    \mathcal{H}_0={\rm diag}\, \left(
  H_0, H_0 \right),
\end{equation}
where $H_0=-\frac{d^2}{dx^2}+1$ is a (shifted for a
constant additive term) free particle Hamiltonian. In other
words, the infinite period limit of the system
$\mathcal{H}(x+{\rm {\bf K}})$ is given by the two copies
of the free non-relativistic particle. As we have seen, the
infinite period limit applied to the integrals of the
self-isospectral system  $\mathcal{H}(x)$ produces
corresponding integrals of the self-isospectral PT system
(\ref{HPTself}). The infinite period limit of the integrals
of the system $\mathcal{H}(x+{\rm {\bf K}})$ may easily be
obtained just by taking a limit $x\rightarrow \infty$ of
the integrals of the self-isospectral PT system
(\ref{HPTself}). For nontrivial local integrals we find
\begin{equation}\label{H0integrals1}
    S_1(x+{\rm {\bf K}})\rightarrow
    -i\frac{d}{dx}\sigma_2-\mathcal{C}_{2\tau}\sigma_1\equiv
s_1\,,\qquad
    S_2(x+{\rm {\bf K}})\rightarrow s_2=i\sigma_3
    s_1\,,\quad
\end{equation}
\begin{equation}\label{H0integrals2}
     Q_a(x+{\rm {\bf K}})\rightarrow
    (-1)^{a+1}\sigma_a\cdot \mathcal{H}_0\,,\quad
    L_1(x+{\rm {\bf K}})\rightarrow -i\frac{d}{dx}\cdot
    \mathcal{H}_0\equiv \ell_1\,,\,\,\,\,
    L_2(x+{\rm {\bf K}})\rightarrow
    \ell_2=\sigma_3\ell_1\,.
\end{equation}
The obtained operators are the integrals of motion for the
trivial free particle supersymmetric system (\ref{calH0}).
They correspond to the obvious integrals $\sigma_a$, and to
the products of them with  $-i\frac{d}{dx}$ and
$\mathcal{H}_0$. System (\ref{calH0}) is intertwined with
the self-isospectral PT system (\ref{HPTself}) by the
infinite period limit of the operator (\ref{Ddiag}),
$
    \hat{\mathcal{D}}\rightarrow {\rm diag}\,\left(
  A_\tau, A_{-\tau}
\right)\equiv \mathcal{D}_\infty,$
$D_\infty\mathcal{H}_0=\mathcal{H}_{PT}D_\infty,$
$\mathcal{H}_0
D_\infty^\dagger=D_\infty^\dagger\mathcal{H}_{PT}.$
If
$J_0$ is some integral for $\mathcal{H}_0$, then $D_\infty
J_0 \mathcal{H}_0 D_\infty^\dagger= D_\infty J_0
 D_\infty^\dagger\mathcal{H}_{PT}$.
Taking into account (\ref{H0integrals1}) and
(\ref{H0integrals2}), the nontrivial local integrals
$S_{PT,a}$, $Q_{PT,a}$ and $L_{PT,a}$ of the
self-isospectral PT system (\ref{HPTself}) may be treated
as a Darboux dressed form of the integrals for the free
particle system $\mathcal{H}_0$, namely, of
 $s_a$, $\sigma_a$ and $-i\mathcal{I}_a\frac{d}{dx}$,
where $\mathcal{I}_1=\mathds{1}$ and
$\mathcal{I}_2=\sigma_3$.

It is interesting to note that the first order integral of
$\mathcal{H}_0$, for instance, $s_1$, may also be treated
as a Hamiltonian of a free relativistic Dirac particle of
mass $\mathcal{C}_{2\tau}$. Then its Darboux dressed form
is a non-periodic BdG Hamiltonian
\begin{equation}\label{BdGPT}
    S_{PT,1}=-i\frac{d}{dx}\sigma_2-\Delta_\tau(x)\sigma_1,
\end{equation}
see Eqs. (\ref{S1limit}) and (\ref{Deltalim}). Comparing
(\ref{BdGPT}) with the structure of $s_1$ in
(\ref{H0integrals1}), we see that  a gap function
$\Delta_\tau(x)$ is effectively a Darboux dressed form of a
free Dirac particle's mass $\mathcal{C}_{2\tau}$. The
periodic BdG Hamiltonian $H_{BdG}=S_1$ may be treated then
as a periodized form of (\ref{BdGPT}), like the Lam\'e
Hamiltonian may be considered as a periodized form of the
PT Hamiltonian, see \cite{SaxBish}. It is worth to stress,
however, that a reconstruction of a crystal structure on
the basis of a non-periodic kink-antikink system is not
direct and free of ambiguities\,: in the previous section
we already noted that two different basic functions of the
shift parameter in the self-isospectral Lam\'e and
associated BdG systems correspond to the same function in
the non-periodic case.

Another interesting observation can be made on a genesis of
the non-local integrals (\ref{SQtilde}). For
self-isospectral Lam\'e and PT systems, the reflection
operator $\mathcal{R}$ and $\sigma_a$, $a=1,2$, are not
integrals of motion, but the  product of any two of these
three operators is an integral of motion. For
supersymmetric free particle system (\ref{calH0}), however,
each of these three operators is an integral of motion. One
finds then  that the infinite period limit of the integral
$\sigma_3\tilde{Q}$, $ \sigma_3\tilde{Q}\rightarrow {\rm
diag}\,\left(
  \mathcal{R}Y_\tau, \mathcal{R}Y_{-\tau}
\right)\equiv\sigma_3\tilde{Q}_{PT}$
is exactly a Darboux
dressed form of the reflection operator $\mathcal{R}$,
$D_\infty \mathcal{R}
D_\infty^\dagger=\sigma_3\tilde{Q}_{PT}$. Or,
alternatively, an integral $\tilde{Q}_{PT}$ for the
self-isospectral PT system is a dressed form of the
nonlocal diagonal integral $\mathcal{R}\sigma_3$. An
analogous relation exists also for the infinite period
limit of another nonlocal diagonal integral from
(\ref{SQtilde}), $D_\infty (-i\mathcal{R}\sigma_2 s_1)
D_\infty^\dagger=\tilde{S}_{PT}\cdot \mathcal{H}_{PT}$,
where $\tilde{S}_{PT}=diag\,
(\mathcal{R}X_\tau,\mathcal{R}X_{-\tau})$. \vskip0.1cm

We conclude that the described Darboux dressing structure
of the self-isospectral PT system, observed earlier in
\cite{PRDPT}, originates from, and is explained
by the properties of the self-isospectral
periodic one-gap Lam\'e system.

\section{Discussion and outlook}

To conclude, let us discuss the obtained results from the
physics perspective and potential applications and
generalizations.

Usual supersymmetric structure of the  kink-antikink as
well as of the   kink crystalline phases of the GN model is
known for about twenty years. However, such a structure
with  the first order supercharges and  $\Z_2$ grading
provided by the diagonal Pauli matrix  does not explain or
reflect a peculiar, finite-gap  nature of the corresponding
solutions. It does not reflect either a  restoration of the
discrete chiral symmetry at zero value of the bare mass in
the GN model, when the kink-antikink crystalline condensate
transforms into the kink crystal. The both aspects are
explained by the exotic nonlinear supersymmetric structure
we revealed here. The finite-gap nature is reflected by the
Lax integral incorporated into a nonlinear supersymmetric
structure alongside with the first and second order
supercharges. A restoration of the discrete chiral
symmetry, on the other hand, is reflected by structural
changes that happen in nonlinear supersymmetry at the half
period shift of Lam\'e superpartner systems, when a central
gap in the spectrum of the associated BdG system
disappears. We showed that the first order BdG
system~\footnote{It is this first order system that really
describes the corresponding crystalline phases in the GN
model while the second order Lam\'e system is related to it
as the Klein-Gordon equation is related to the Dirac
equation.} has its own supersymmetry, which can be revealed
only with the help of the nonlocal grading operators
investigated in Section 6. The disappearance of the middle
gap in the BdG spectrum is accompanied by emergence of the
new, nontrivial second order integral of motion in the
first order system (while the BdG Hamiltonian has no such
integral in the kink-antikink crystalline phase).

The aspects related to the infinite period limit we
investigated in sections 8 and 9 may  be useful for
understanding of some puzzles  related  to a computation of
the Witten index in some supersymmetric field theories when
a system is put in a  periodized box \cite{Smilga}.

Recently, perfect Klein tunneling in carbon nanostructures
was explained in \cite{JP1} by  unusual supersymmetric
structure with the first order matrix Hamiltonian. We
believe that the supersymmetry we investigated here,
particularly in Section 7,  may also be useful in the study
of other phenomena in graphene,  where the dynamics of
charges is governed by the effective first order Dirac
Hamiltonian.

It would be interesting to clarify whether the twisted kink
crystal of the GN model with continuous chiral symmetry,
that was found in \cite{BasDun1,BasDun2}, could be obtained
by supersymmetric constructions similar to those from
section 3.

We treated $\lambda$ that appears in the structure of the
second order intertwining operator
$\mathcal{B}(x;\tau,\lambda)$ of a general form
(\ref{Acal}) as a kind of a virtual shift parameter. One
could extend the picture by reinterpreting Eqs.
(\ref{Htlam}) and (\ref{AcalH}) as intertwining relations
for three Lam\'e systems, $H(x+\tau_1)$, $H(x+\tau_2)$ and
$H(x+\tau_3)$, where $\tau_1=\tau$, $\tau_2=\tau+2\lambda$
and $\tau_3=-\tau$. Then we would get an extended
self-isospectral system of three super-partner Lam\'e
Hamiltonians. Employing relation of a form
(\ref{tau1tau2}), one could further extend the picture to
obtain a self-isospectral system with $n>3$ superpartners
$H(x+\tau_1),\ldots, H(x+\tau_n)$. When the shift
parameters are such that $\tau_n=\tau_1$, the corresponding
intertwining operator of order $n$ would reduce to an
integral for the system $H(x+\tau_1)$. It is in such a way
we identified, in fact,  the third order Lax operator
$\mathcal{P}(x+\tau)$ for the system $H(x+\tau)$. The
interesting questions that arise are then\,: what is a
complete set of integrals and what kind of supersymmetry we
get for such an $n$-component self-isospectral system?
Particularly, what is the nature of the above-mentioned
integral of motion of the order $n$ for $n>3$? What is a
relation of such extended supersymmetric systems with the
GN model and  what physics could be associated with them?

\vskip0.2cm

 \noindent \textbf{Acknowledgements.}
The work of MSP has been partially supported by
 FONDECYT Grant 1095027, Chile and  by Spanish Ministerio de
 Educaci\'on under Project
SAB2009-0181 (sabbatical grant). LMN has been partially
supported by the Spanish Ministerio de Ciencia e
Innovaci\'on (Project MTM2009-10751) and Junta de Castilla
y Le\'on (Excellence Project GR224). MSP and AA thank
Physics Department of Valladolid University for
hospitality.

\section*{Appendix A:\,  Jacobi elliptic functions}\label{ap1}
\renewcommand{\theequation}{A.\arabic{equation}}
\setcounter{equation}{0}

We summarize here some  properties and relations for Jacobi
elliptic and related functions. For details, see, e.~g.,
\cite{WW,Akhi}. \vskip0.05cm

In notations for these functions we suppress a dependence
on a modular parameter $0<k<1$, $\sn\,x=\sn(x|k)$, etc.,
when this does not lead to ambiguities.  On the other hand,
a dependence on a complementary modulus parameter $0<k'<1$,
$k'=(1-k^2)^{1/2}$, is indicated explicitly. We use
Glaisher's notation for inverse quantities and quotients
of Jacobi elliptic functions, $\nd\,x=1/\dn\,x$,
$\ns\,x=1/\sn\,x$, $\nc\,x=1/\cn\,x$,
$\sc\,x=\sn\,x/\cn\,x$, etc. \vskip0.05cm

 The basic Jacobi elliptic functions  are the
doubly-periodic meromorphic functions $\sn\,u$, $\cn\,u$
and $\dn\,u$, whose periods are ($4{\rm {\bf K}}$, $2i{\rm
{\bf K}}'$), ($4{\rm {\bf K}}$, $2{\rm {\bf K}}+2i{\rm {\bf
K}}'$) and ($2{\rm {\bf K}}$, $4i{\rm {\bf K}}'$),
respectively. $\sn\,u$ is an odd function, while $\cn\,u$
and $\dn\,u$ are even functions, which are related by
identities $\sn^2 u + \cn^2 u=1,$ $\dn^2 u+k^2\sn^2 u=1,$
$k^2\cn^2 u +k'{}^2=\dn^2 u,$ $k'{}^2\sn^2 u + \cn^2
u=\dn^2 u,$ and whose derivatives are  $
    \frac{d}{du}\,\sn\,u=\cn\,u\,\dn\,u,
$ $    \frac{d}{du}\,\cn\,u=-\sn\,u\,\dn\,u, $ $
    \frac{d}{du}\,\dn\,u=-k^2\sn\,u\,\cn\,u\,.
$
 They have simple zeros and poles
 at
\begin{equation}\label{zeros}
     \sn\,u\,:\,\,0,\,2{\rm {\bf K}}\,;\qquad
    \cn\,u\,:\,\, {\rm {\bf K}},\,-{\rm {\bf K}}\,;\qquad
    \dn\,u\,:\,\,{\rm {\bf K}}+i{\rm {\bf K}}',
    \,{\rm {\bf K}}-i{\rm {\bf K}}'\,,
\end{equation}
\begin{equation}\label{poles}
    \sn\,u\,,\,\,\cn\,u\,:\,\,i{\rm {\bf K}}',\,2{\rm {\bf K}}+
    2i{\rm {\bf K}}'\,;\qquad
    \dn\,u\,:\,\,i{\rm {\bf K}}',\,-i{\rm {\bf K}}'\,,
\end{equation}
respectively, modulo periods. Here
\begin{equation}\label{KK'}
    {\rm {\bf K}}={\rm {\bf K}}(k)=\int_0^1\frac{dx}{\sqrt{(1-x^2)(1-k^2x^2)}}
\end{equation}
is a complete elliptic integral of the first kind, and $
{\rm {\bf K}}'={\rm {\bf K}}(k')$ is a complementary
integral, which are monotonic functions of $k$ in the
interval $0<k<1$\,: $d{\rm {\bf K}}/dk>0$, $d{\rm {\bf
K}}'/dk<0$. In the limit cases $k=0$ and $k=1$, elliptic
functions transform into simply-periodic functions in a
complex plane,
\begin{eqnarray}\label{limit1}
    &k=0,\,k'=1\,:\quad
    {\rm {\bf K}}=\frac{1}{2}\pi,\,{\rm {\bf K}}'=\infty\,,\quad
    \sn\,u=\sin u,\,\,\, \cn\, u=\cos u,\,\,\,\dn\,u=1,&\\
    \label{limit2}
    &k=1,\,k'=0\,:\quad
    {\rm {\bf K}}=\infty,\,{\rm {\bf K}}'=\frac{1}{2}\pi\,,\quad
    \sn\,u=\tanh u,\,\,\, \cn\, u=\dn\,u=\frac{1}{\cosh
    u}.&
\end{eqnarray}
The addition formulae are
\begin{eqnarray}
    {\rm s}_+=\frac{1}{\mu}({\rm s}_u{\rm c}_v
    {\rm d}_v+{\rm s}_v{\rm c}_u{\rm d}_u),\quad
    {\rm c}_+=\frac{1}{\mu}({\rm c}_u{\rm c}_v
    -{\rm s}_u{\rm s}_v{\rm d}_u{\rm d}_u),\quad
    {\rm d}_+=\frac{1}{\mu}({\rm d}_u{\rm d}_v
    -k^2{\rm s}_u{\rm s}_v{\rm c}_u{\rm c}_u),
    \label{addJs}
\end{eqnarray}
where ${\rm s}_+=\sn\,(u+v)$, ${\rm s}_u=\sn\, u$, ${\rm
s}_v=\sn\, v$, ${\rm c}_+=\cn\, (u+v)$, ${\rm
d}_+=\dn\,(u+v)$, etc., and $
    \mu=1-k^2\sn^2u\,\sn^2v\,.
$ The Jacobi's imaginary transformation is
\begin{equation}\label{Jacim}
    \sn(iu|k)=i\sn(u|k')\nc(u|k'),\quad
    \cn(iu|k)=\nc(u|k'),\quad
    \dn(iu|k)=\dn(u|k')\nc(u|k')\,.
\end{equation}
{}From addition formulae and (\ref{Jacim}), one finds  some
displacement properties of Jacobi elliptic functions shown
in Table \ref{T2}.
\begin{table}[ht]
\caption{Displacement properties of Jacobi elliptic
functions} \label{T2}
\begin{center}
\begin{tabular}{|c||c|c|c|c|c|c|}\hline
$u$ & $u+{\rm {\bf K}}$ & $u+i{\rm {\bf K}}'$ & $u+{\rm
{\bf K}}+i{\rm {\bf K}}'$ & $u+2{\rm {\bf K}}$ & $u+2i{\rm
{\bf K}}'$ & $u+2({\rm {\bf K}}+i{\rm {\bf K}}')$
\\\hline\hline
$\sn\, u$  & $\cn\, u\,\nd\,u$ & $\frac{1}{k}\,\ns\,u$ &
$\frac{1}{k}\,\dn\,u\,\nc\,u$ & $-\sn\,u$ & $\sn\,u$ &
$-\sn\,u$
\\[1pt]\hline
$\cn\,u$  & $-k'\sn\,u\,\nd\,u$ &
$-i\frac{1}{k}\,\dn\,u\,\ns\,u $ & $- i\frac{k'}{k}\,\nc\,u
$ & $-\cn\,u$ & $-\cn\,u$ & $\cn\,u$
 \\[1pt]\hline
$\dn\,u$  & $k'\nd\,u$ & $-i\cn\,u\,\ns\,u $ & $ik'
\sn\,u\,\nc\,u $ & $\dn\,u$ & $-\dn\,u$ & $-\dn\,u$
 \\[1pt]\hline
\end{tabular}
\end{center}
\end{table}

\section*{Appendix B:\,  Jacobi Zeta, Theta and Eta functions}\label{bp1}
\renewcommand{\theequation}{B.\arabic{equation}}
\setcounter{equation}{0}

The complete elliptic integral of the second kind is
defined by
\begin{equation}\label{EK}
    {\rm \bf{E}}={\rm \bf{E}}(k)=\int_0^1\sqrt{\frac{1-k^2x^2}{1-x^2}}\,dx
    \,.
\end{equation}
It is a monotonically decreasing function, $d{\rm
\bf{E}}/dk<0$. The complete elliptic integrals ${\rm {\bf
K}}={\rm \bf{K}}(k)$  and ${\rm \bf{E}}={\rm \bf{E}}(k)$
satisfy the first order differential equations $
    \frac{d{\rm {\bf K}}}{dk}=\frac{{\rm \bf{E}}-
    k'{}^2{\rm {\bf K}}}{kk'{}^2},
$
$
    \frac{d{\rm \bf{E}}}{dk}=\frac{{\rm \bf{E}}-{\rm {\bf
    K}}}{k},$
from which an inequality  $k'{}^2<{\rm {\bf E}}/{\rm {\bf
K}}<1 $ and  the Legendre's relation
$
    {\rm \bf{E}}{\rm {\bf K}}'+{\rm \bf{E}}'{\rm {\bf K}}-
    {\rm {\bf K}}{\rm {\bf K}}'=\frac{1}{2}\pi$
may be deduced, where ${\rm \bf{E}}'={\rm \bf{E}}(k')$ is a
complementary integral of the second kind.

The incomplete elliptic integral of the second kind is
defined as
\begin{equation}\label{E(u)}
    {\rm E}(u)=\int_0^u\dn^2u\, du\,,
\end{equation}
in terms of which ${\rm \bf{E}}={\rm E}({\rm \bf{K}})$.
This is an odd analytic function of $u$, regular save for
simple poles of residue $+1$ at the points $2nK+(2m+1)iK'$.
Function ${\rm E}(u)$ is not an elliptic function. It
possesses the properties of pseudo-periodicity,
$
    {\rm E}(u+2{\rm {\bf K}})-{\rm E}(u)={\rm E}(2{\rm {\bf K}})=2{\rm
    \bf{E}},$
$
    {\rm E}(u+2i{\rm {\bf K}}')-{\rm E}(u)={\rm E}(2i{\rm {\bf
    K}}'),
$ where in the first relation the second equality is
obtained by putting $u=-{\rm {\bf K}}$.

In terms of ${\rm E}(u)$, a simply-periodic Jacobi Zeta
function is defined,
\begin{equation}\label{Zdes}
    Z(u)={\rm E}(u)-\frac{{\rm \bf{E}}}{{\rm {\bf
    K}}}\,u\,,
\end{equation}
which satisfies relations
$
    \frac{d{\rm Z}(u)}{du}=\dn^2 u-
    \frac{{\rm \bf{E}}}{{\rm {\bf K}}}\,,
$
and
\begin{eqnarray}\label{Z1}
    &{\rm Z}(u+2{\rm {\bf K}})={\rm Z}(u),\,\,
    {\rm Z}(u+2i{\rm {\bf K}}')={\rm Z}(u)-i\frac{\pi}{{\rm {\bf K}}},
    \,\,
    {\rm Z}(-u)=-{\rm Z}(u),\,\,
    {\rm Z}({\rm {\bf K}}-u)=-{\rm Z}({\rm {\bf K}}+u),\quad &\\
    \label{ZKK'}
    &{\rm Z}(0)={\rm Z}({\rm {\bf K}})=0,\qquad
    {\rm Z}({\rm {\bf K}}+i{\rm {\bf K}}')=
    -i\frac{\pi}{2{\rm {\bf K}}}.&
\end{eqnarray}
Zeta function satisfies an addition formula
\begin{equation}\label{Zadd}
    {\rm Z}(u+v)={\rm Z}(u)+{\rm
Z}(v)-k^2\sn\,u\,\sn\,v\,\sn\,(u+v)\,,
\end{equation}
and obeys Jacobi's imaginary transformation
\begin{equation}\label{Zim}
    i{\rm Z}(iu|k)={\rm Z}(u|k')+\frac{\pi u}{2{\rm {\bf K}}{\rm {\bf K}}'}
    -\dn(u|k')\sc(u|k')\,,
\end{equation}
from which one finds $
    {\rm Z}(u+i{\rm {\bf K}}')={\rm
Z}(u)+\ns\,u\,\cn\, u\,\dn\,u\,-i\frac{\pi}{2{\rm {\bf
K}}}. $ For the limit values of the modular parameter,
$k=0$ and $k=1$, we have
\begin{equation}\label{Z2}
    {\rm Z}(u|0)=0\,,\qquad
    {\rm Z}(u|1)=\tanh u\,.
\end{equation}
In terms of ${\rm Z}(u)={\rm Z}(u|k)$, the Jacobi Theta
function $\Theta(u|k)$ is defined as
\begin{equation}\label{Theta}
    \Theta(u)=\Theta(0)\exp\left(\int_0^u Z(u)\,du\right)\,.
\end{equation}
 This is an even, $\Theta(-u)=\Theta(u)$, integral periodic
function of period $2{\rm {\bf K}}$, whose only zeros are
simple ones at the points of the set $2n{\rm {\bf
K}}+(2m+1)i{\rm {\bf K}}'$.  It satisfies a relation
$
    \Theta(u+2i{\rm {\bf K}}')=-\frac{1}{q}\,
    \exp\left(-i\frac{\pi}{{\rm {\bf K}}}u\right)
    \Theta(u),
$
where
$
    q=\exp(- \pi {\rm {\bf K}}'/{\rm {\bf K}}).
$ Notice  that sometimes Jacobi's Theta function is defined
by the Fourier series,
\begin{equation}\label{Z0}
    \Theta(u|k)=\vartheta_4(v)\,,\quad
    \vartheta_4(z)=1+2\sum_{n=1}^\infty
    (-1)^n q^{n^2}\cos(2nz)\,,
    \quad
    v=\frac{\pi u}{2K}\,.
\end{equation}
Then ${\rm Z}$  function can be defined by the logarithmic
derivative,
\begin{equation}\label{ZTheta}
{\rm Z}(u)=\frac{d}{du}\ln \Theta(u)\,.
\end{equation}
In correspondence with definition (\ref{Z0}), a constant in
(\ref{Theta}) is fixed as
$
    \Theta(0)=\sqrt{\frac{2{\rm {\bf K}}k{}\,'}{\pi}}.
$

Jacobi Eta function ${\rm H}(u)$ is defined in terms of the
Theta function,
\begin{equation}\label{Eta}
    {\rm H}(u)=-iq^{1/4}\exp\left(i\frac{\pi u}{2{\rm {\bf K}}}\right)
    \Theta(u+i{\rm {\bf K}}')\,.
\end{equation}
This is an odd, ${\rm H}(-u)=-{\rm H}(u)$, integral
periodic  function of period $4{\rm {\bf K}}$, which
possesses simple zeros at the points of the set $2n{\rm
{\bf K}}+2mi{\rm {\bf K}}'$. Some properties of the Eta and
Theta functions are summarized in Table \ref{T3},
\begin{table}[ht]
\caption{Parity and some displacement properties of Jacobi
$\Theta$ and ${\rm H}$ functions} \label{T3}
\begin{center}
\begin{tabular}{|c||c|c|c|c|c|c|}\hline
$u$ & $-u$ & $u+2{\rm {\bf K}}$ & $u+i{\rm {\bf K}}'$ &
$u+2i{\rm {\bf K}}'$ & $u+{\rm {\bf K}}+i{\rm {\bf K}}'$ &
$u+2{\rm {\bf K}}+2i{\rm {\bf K}}'$
\\\hline\hline
$\Theta(u)$ & $\Theta(u)$ & $\Theta(u)$ & $iM(u){\rm H}(u)$
& $-N(u)\Theta(u)$ & $M(u){\rm H}(u+{\rm {\bf K}})$ &
$-N(u)\Theta(u)$
\\[1pt]\hline
${\rm H}(u)$ & $-{\rm H}(u)$ & $-{\rm H}(u)$ &
$iM(u)\Theta(u)$ & $-N(u){\rm H}(u)$ & $M(u)\Theta(u+{\rm
{\bf K}})$ & $N(u){\rm H}(u)$
 \\[1pt]\hline
\end{tabular}
\end{center}
\end{table}
where
$
    M(u)=\exp\left(-i\frac{\pi
    u}{2{\rm {\bf K}}}\right)q^{-1/4},$
$
    N(u)=\exp\left(-i\frac{\pi
    u}{{\rm {\bf K}}}\right)q^{-1}\,.
$ For particular values of the argument we also have
$
    {\rm H}'(0)=\frac{\pi}{2{\rm {\bf K}}}{\rm
    H}({\rm {\bf K}})\Theta(0)\Theta({\rm {\bf K}}),$
$
    \Theta({\rm {\bf K}})=\sqrt{\frac{2{\rm {\bf
    K}}}{\pi}},
$
$
    {\rm H}({\rm {\bf K}})=\sqrt{\frac{2k{\rm {\bf K}}}{\pi}}.
$ Jacobi Theta function satisfies a kind of addition
theorem,
\begin{equation}\label{ThetaAdd}
    \Theta(u+v)\Theta(u-v)\Theta^2(0)=\Theta^2(u)\Theta^2(v)-
    {\rm H}^2(u){\rm H}^2(v)\,.
\end{equation}
The basic Jacobi elliptic functions may be represented in
terms of $\Theta$ and ${\rm H}$ functions,
\begin{equation}\label{elThH}
    \sn\,u=\frac{{\rm H}(u)}{\Theta(u)}\cdot
    \frac{\Theta(0)}{{\rm H}'(0)}\,,\qquad
    \cn\,u=\frac{{\rm H}(u+{\rm {\bf K}})}{\Theta(u)}\cdot
    \frac{\Theta(0)}{{\rm H}({\rm {\bf K}})}\,,\qquad
    \dn\,u=\frac{\Theta(u+{\rm {\bf K}})}{\Theta(u)}\cdot
    \frac{\Theta(0)}{\Theta({\rm {\bf K}})}\,.
\end{equation}

Under the complex conjugation, all the Jacobi elliptic
functions as well as ${\rm H}$, $\Theta$ and ${\rm Z}$
satisfy a relation
$
    (f(z))^*=f(z^*).
$


\end{document}